\documentclass[preprint2,10pt]{aastex}

\begin{document}

\title{Morphological Classification of Galaxies by Shapelet Decomposition in the Sloan Digital Sky Survey}
\author{Brandon C. Kelly}
\affil{Steward Obvservatory, Tucson, AZ 85721-0065}
\email{bkelly@as.arizona.edu}
\author{Timothy A. McKay}
\affil{Physics Department, University of Michigan, Ann Arbor, MI 48109-1090}
\email{tamckay@umich.edu}

\begin{abstract}

We describe application of the `shapelet' linear decomposition of galaxy images to morphological classification using images of $\sim$ 3000 galaxies from the Sloan Digital Sky Survey.  After decomposing the galaxies we perform a principal component analysis to reduce the number of dimensions of the shapelet space to nine.  We find that each of these nine principal components contains unique morphological information, and give a description of each principal component's contribution to a galaxy's morphology.  We find that galaxies of differing Hubble type separate cleanly in the shapelet space.  We apply a Gaussian mixture model to the 9-dimensional space spanned by the principal components and use the results as a basis for classification.  Using the mixture model, we separate galaxies into seven classes and give a description of each class's physical and morphological properties.  We find that several of the mixture model classes correlate well with the traditional Hubble types both in their morphology and their physical parameters (e.g., color, velocity dispersions, etc.). In addition, we find an additional class of late-type morphology but with high velocity dispersions and very blue color; most of these galaxies exhibit post-starburst activity.  This method provides an objective and quantitative alternative to traditional and subjective visual classification.

\end{abstract}

\keywords{methods : data analysis --- methods : statistical --- techniques : image processing --- galaxies : fundamental parameters --- galaxies : statistics}

\section{INTRODUCTION}

\label{intro}

The first widely accepted scheme for morphological classification of galaxies was the Hubble tuning fork \citep{hub36}.  Although Hubble's system has proven useful as a framework for discussion, and has provided insights into the relationships between galaxy morphology, galactic evolution, stellar populations, and other physical quantities \citep{bergh98,elm98}, it is lacking in several respects.  Most importantly, the Hubble scheme is subjective in the sense that morphological differences between classes are not robustly defined, and it relies on manual classification by experts.  Also, there are many galaxies that do not fit into the Hubble sequence, such as enormous cD galaxies \citep{mat64, mor65}, H I gas-poor anemic galaxies \citep{bergh76}, and dusty amorphous galaxies \citep{san79}.  Often such galaxies are `pigeon-holed' into inappropriate classifications, resulting from the Hubble system's lack of flexibility and use of discrete classes. In addition, many low luminosity dwarf galaxies do not easily fit into Hubble's scheme \citep{elm98}, nor do those known as peculiar galaxies \citep{arp66}.  Hubble's classification scheme does not provide an adequate framework for high redshift ($z > 0.3$) galaxies \citep{bergh01, abr96a} or those in the cores of rich clusters \citep{koop98,abr94}.  This is not surprising, as Hubble based his system on the morphologies of galaxies in the nearby universe.  Galaxies at high redshift are seen at significantly earlier stages in their evolution, and are more likely to undergo mergers and collisions that distort their morphology.  In addition, the elliptical sequence EO--E7 of the tuning fork is heavily dependent on the orientation of the galaxy, rather than its intrinsic morphology.

Motivated by the ambiguities in Hubble's classification framework, there have been many additions and modifications to it.  In an attempt to make the Hubble sequence more complete, \citet{devauc59} introduced finer divisions regarding barred spirals and differentiation between those galaxies showing rings and those that do not.  However, the addition of these parameters has not been conclusively tied to physically significant differences among galaxies \citep{kor82}.  A twelve-stage classification sequence for spiral arms was developed by \citet{elm82}, where the classifications range from patchy `flocculent' arms to well defined `grand design' arms.  It has been shown that arm type and Hubble type appear to be unrelated \citep{bergh98}.  The addition of luminosity classes modeled after stellar luminosity classifications was given by \citet{bergh60}.  A simpler system of galaxy classification based on the central concentration of light was developed by \citet{mor58}, and has proven to be useful for classifying galaxies at large redshifts and in the cores of rich clusters \citep{bergh98}.

Most of the recent work on galaxy classification has been in the pursuit of an automated and objective classification system.  It has been shown that manual classification of images with limited dynamical range and poor resolution leads to a disappointingly large difference among human classifiers \citep{naim95, abr96b}.  This discrepancy is most pronounced at high redshifts, where the traditional Hubble sequence begins to break down. Quantitative conclusions of how a galaxy's morphology is related to its physical parameters and evolution have been hindered by the qualitative and environment-dependent nature of the Hubble classifications.  Many physical properties of galaxies have been shown to be correlated \citep[e.g.,][]{blan03} and one would desire a quantitative description of how a galaxy's morphology fits in; i.e., a morphological classification scheme should probe the important physical properties of galaxies.  In addition, the advent of enormous astronomical databases necessitates automatic classification.  For example, the Sloan Digital Sky Survey \citep[SDSS,][]{york00} is expected to catalog the spectra for roughly one million galaxies, making manual classification impractical.

Many methods of automatic and objective classification have been proposed.  A method which utilizes the central concentration index of a galaxy was proposed by \citet{doi93}, and \citet{abr94,abr96b} have developed a promising system based on a galaxy's central concentration of light and asymmetry.  \citet{cons03} has taken this a step further through the addition of a `clumpiness' parameter, and relates the central concentration, asymmetry, and clumpiness parameters to the underlying physical processes and history.  Much of the work toward automatic galaxy classification has been in the use of artificial neural networks \citep[e.g.,][]{god02,ode95}).  Neural networks are powerful learning tools, where the central idea is to extract linear combinations of the inputs as derived features, and then model the target as a nonlinear function of these features \citep{hast01}.  Neural networks have met with approximately the same success as visual classification \citep[e.g.,][]{ball03}, but have relied on the training set for classification; neural networks can replicate visual classification, but not create a classification framework.  Additional techniques include principal component analysis \citep{frei99} and fractal dimensions \citep{tha00}.  Those classification schemes that are able to separate galaxies cleanly often utilize only $\sim 100-200$ galaxies, resulting in a fairly small number of galaxies in each class and making comparison difficult.  In addition, a complete quantitative description of galaxy morphology is still lacking, as most systems utilize only two or three parameters \citep[e.g.,][]{doi93,abr00}.  Techniques involving Fourier analysis also hold promise \citep{ode02,tri98}, but rely on analyzing azimuthal profiles of a galaxy, rather than analyzing the full 2-dimensional image of the galaxy.

One promising approach toward a quantitative, complete, and automatic system of morphological classification is to linearly decompose a galaxy into `shapelets' \citep{ref03a}.  Shapelets are Gaussian-weighted Hermite polynomials which happen to be the eigenstates of the quantum harmonic oscillator Hamiltonian.  The have been shown to be useful in weak gravitational lensing measurements \citep{chang02,ref03b} and for image simulation \citep{mas03}.  They form an ideal framework in which to classify galaxies as they are mathematically well defined and understood, compact, form an orthonormal basis, and use all of the shape information in a galaxy.  Shapelets are preferred over other techniques that utilize a basis expansion, such as wavelets \citep{far92} or curvelets and ridgelets \citep{star03}.  These techniques rely on translated and dilated versions of a single shape (e.g., a Gaussian wave packet in the case of the Morlet wavelet), while shapelets form a complete set of varying shapes with the same scale.  Because the shapelet decomposition maps a 2-dimensional galaxy image onto a vector in a multi-dimensional space, galaxy classification amounts to finding structure in the shapelet coefficient space.  The shapelet method provides a continuous framework to describe galaxy morphology and classification, rather than several discrete bins as in traditional classification methods.

Because differences in resolution contribute to differences in observed morphology we choose to artificially redshift all galaxies and smooth them to the same physical scale.  This assures that galaxies are resolved on the same physical scale, rather than on the same angular scale, ensuring that the resolution is independent of galaxy distance or size. Other techniques have used a similar method \citep{abr96b, cons03}, however we do not know of any frameworks that measure the morphology of galaxies smoothed to the same {\em physical} scale.  The technique is fully automatic, allowing efficient analysis of galaxy images.  We apply the method to data from the SDSS, using both galaxies previously classified according to Hubble type and previously unclassified galaxies.

\section{SHAPELETS}

\subsection{Cartesian Shapelets}

\label{cart}

Much of the shapelet formalism can be found in \citet{ref03a}. We first construct the dimensionless basis function in 1-dimension:
\begin{equation}
\phi_n(x) \equiv \left[ 2^n \pi^{\frac{1}{2}} n! \right]^{-\frac{1}{2}} H_n(x) e^{-\frac{x^2}{2}},
\label{eq01}
\end{equation}
where $n$ is a non-negative integer and $H_n(x)$ a Hermite polynomial of order $n$.  In practice, we use the dimensional basis functions, which are easily described from the dimensionless set by
\begin{equation}
B_n(x; \gamma) \equiv \gamma^{-\frac{1}{2}} \phi_n( \gamma^{-1} x),
\label{eq02}
\end{equation}
where $\gamma$ is the characteristic scale of the object to be analyzed.  These functions are orthonormal, i.e.
\begin{equation}
\int_{-\infty}^{\infty} B_n(x ; \gamma) B_m(x ; \gamma) dx = \delta_{nm}.
\label{eq03}
\end{equation}

Because the Cartesian shapelets are separable in $x$, we can easily construct the 2-dimensional basis functions from the 1-dimensional:
\begin{equation}
\phi_{\bf n}({\bf x}) \equiv \phi_{n_1}(x_1) \phi_{n_2}(x_2), 
\label{eq04}
\end{equation}
where ${\bf x}=(x_1,x_2)$ and ${\bf n}=(n_1,n_2)$.  The dimensional basis functions become
\begin{equation}
B_{\bf n}({\bf x},\gamma) \equiv \gamma^{-1} \phi_{\bf n}(\gamma^{-1} {\bf x}), 
\label{eq05}
\end{equation}
and are orthonormal.

\begin{figure}
\begin{center}
\scalebox{0.8}{\rotatebox{90}{\plotone{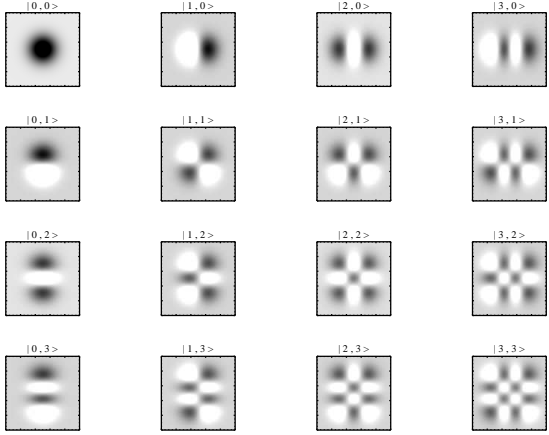}}}
\caption{Plots of the first several 2-dimensional Cartesian shapelets $| n_1 , n_2 \rangle$. Dark areas are positive, light are negative. \label{fig01}}
\end{center}
\end{figure}

These 2-dimensional shapelets form a complete orthonormal basis for smooth and integrable functions of two variables, allowing us to decompose any well-behaved two-dimensional image $f({\bf x})$ into a sum of shapelets as
\begin{equation}
f({\bf x}) = \sum_{n_1, n_2 = 0}^{\infty} f_{\bf n} B_{\bf n}({\bf x}; \gamma),
\label{eq06}
\end{equation}
with shapelet coefficients
\begin{equation}
f_{\bf n} = \int f({\bf x}) B_{\bf n}({\bf x}; \gamma) d^2 x,
\label{eq07}
\end{equation}
from the orthonormality property.  The first several 2-dimensional Cartesian shapelets are shown in Figure \ref{fig01}. Dirac notation is a convenient and concise way of representing much of the math of quantum mechanics, and we will often use it in this paper.  In Dirac notation, the $n^{th}$ state is denoted as $|n \rangle $, and has $x$-space representation $\langle x|n \rangle =\phi_n$.

\subsection{Flux and Centroid}

\label{flux}

The flux and center of an astronomical object are two of the most basic quantities that one measures, and computing them is necessary for any classification scheme that relies on linearly decomposing a galaxy into basis functions.  The total flux $F \equiv \int f({\bf x}) d^2 x$ of the galaxy may be computed from the shapelet coefficients as
\begin{equation}
F = \pi^{\frac{1}{2}} \gamma \sum_{n_1,n_2}^{even} 2^{(2-n_1-n_2)/2} \left( \ \stackrel{\displaystyle n_1}{n_1/2} \ \right)^{\frac{1}{2}} \left( \ \stackrel{\displaystyle n_2}{n_2/2} \ \right)^{\frac{1}{2}} f_{n_1 n_2},
\label{eq08}
\end{equation}
where the sum is over even $n_1, n_2$ and the parentheses signify the binomial coefficient.  The centroid of an object $x_{Ci} \equiv \int x_i f({\bf x}) d^2 x /F$ is given by
\begin{eqnarray}
\lefteqn{x_{C1} = \frac{\pi^\frac{1}{2} \gamma^2}{F} \sum_{n_1}^{odd} \sum_{n_2}^{even} (n_1+1)^{\frac{1}{2}} 2^{\frac{1}{2} (2 - n_1 - n_2)} \ \times} \nonumber \\
 & & \left( \ \stackrel{\displaystyle n_1+1}{(n_1+1)/2} \ \right)^{\frac{1}{2}} \left( \ \stackrel{\displaystyle n_2}{n_2/2} \ \right)^{\frac{1}{2}} f_{n_1 n_2},
\label{eq09}
\end{eqnarray}
and similarly for $x_{C2}$.

\subsection{Polar Shapelets}

\label{polar}

\begin{figure}
\begin{center}
\scalebox{0.8}{\rotatebox{90}{\plotone{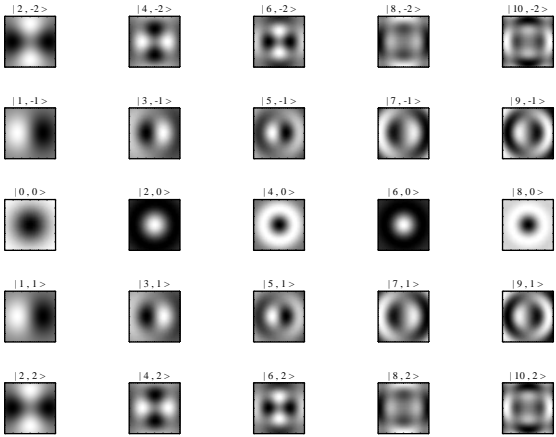}}}
\caption{Plots of the real part of the first several 2-dimensional Polar shapelets $|n, m \rangle$. \label{fig02}}
\end{center}
\end{figure}

It is of practical interest to construct shapelets separable in the polar coordinates $r$ and $\varphi$, as polar coordinates are more appropriate for describing a galaxy's shape.  The polar basis functions are eigenstates of the QHO Hamiltonian and angular momentum simultaneously.  One can show that in $x$-space the polar basis functions $\chi_{n_l,n_r}(r, \varphi) \equiv \langle x | n_l,n_r \rangle$ are given by
\begin{equation}
\chi_{n_l n_r}(r, \varphi) = [\pi n_l ! n_r !]^{-\frac{1}{2}} H_{n_l n_r}(r) e^{-r^2/2} e^{i(n_r - n_l)\varphi},
\label{eq10}
\end{equation}
where $H_{n_l n_r}(r)$ are the `polar Hermite polynomials'.  The energy eigenvalue $n$ and the angular momentum eigenvalue $m$ are related to $n_l$ and $n_r$ by
\begin{eqnarray}
n & = & n_r + n_l \nonumber \\
m & = & n_r - n_l, \ \ \ m = -n, -n+2, \ldots , n - 2, n.
\label{eq11}
\end{eqnarray}
The dimensional polar basis functions are
\begin{equation}
A_{n_l n_r}(r, \varphi; \gamma) = \gamma^{-1} \chi_{n_l n_r}(\gamma^{-1} r, \varphi)
\label{eq12}
\end{equation}
and have the orthonormality property:
\begin{equation}
\int_0^{2 \pi} d \varphi \int_0^{\infty} dr r A_{n_l n_r}(r, \varphi; \gamma) A_{n'_l n'_r}(r, \varphi; \gamma) = \delta_{n_l n'_l} \delta_{n_r n'_r}.
\label{eq13}
\end{equation}
Plots of the first several polar shapelets can also be seen in Figure \ref{fig02}.

We often need to change between Cartesian and polar shapelets, and require a relationship that allows us to do so.  One can show that this is given in the form of the transformation matrix between these two bases,
\begin{eqnarray}
\lefteqn{\langle n_1,n_2 | n_l,n_r \rangle = 2^{-(n_r+n_l)/2} i^{n_r-n_l} \left[ \frac{n_1 ! n_2 !}{n_r ! n_l !} \right]^{\frac{1}{2}} \ \times} \nonumber \\
 & & \delta_{n_1+n_2,n_l+n_r} \sum_{n'_r=0}^{n_r} \sum_{n'_l=0}^{n_l} i^{n'_l - n'_r} \left( \ \stackrel{\displaystyle n_r}{n'_r} \ \right) \left( \ \stackrel{\displaystyle n_l}{n'_l} \ \right) \delta_{n'_r + n'_l, n_1}.
\label{eq14}
\end{eqnarray}
From this equation, we see that only those states with $n_1+n_2=n_l+n_r=n$ are mixed.

\section{THE DATA}

\label{data}

The SDSS \citep{york00} has been producing imaging and spectroscopic surveys of the Northern Galactic Cap over $\pi$ steradians.  A 2.5m telescope at the Apache Point Observatory, Sunspot, New Mexico, observes the sky in five bands \citep[{\it u, g, r, i, z},][]{fuk96,hogg01,smith02} between 3000 and 10000 \AA , using a drift-scanning mosaic CCD camera \citep{gunn98}, which detects objects to a flux limit of $r \sim 22.5$ mags.  The survey, when finished, is expected to spectroscopically observe 900,000 galaxies down to $r_{lim} \approx 17.77$ mags \citep{str02}, 100,000 Luminous Red Galaxies \citep{eis01}, and 100,000 quasars \citep{ric02}.  The spectroscopic follow up uses two digital spectrographs on the same telescope as the imaging camera, and the spectroscopic samples are assigned plates and fibers using an algorithm described by \citet{blan03t}.  The astrometric calibration is described in \citet{pier03}.  Details of the galaxy survey can be found in the galaxy target selection paper \citep{str02}, and other principles of the survey are described in the Early Data Release \citep{sto02}.

We use two samples in this analysis.  We first investigate the shapelet method using the {\it r}-band data for 556 of the 1482 well-resolved galaxies used by \citet[][hereafter Sample 1]{nak03} to estimate the morphology-dependent luminosity function.  We have chosen this group in order to compare the shapelet results with traditional Hubble type, as this catalog contains manual classifications of Hubble type.  The manual classifications make no distinction between spirals with bar structure and spirals without.  Sample 1 contains those galaxies of the \citet{nak03} sample that have redshift $z < 0.07$ and a PSF width of less than 2.0 kpc when projected onto the plane of the galaxy.  This allows us to smooth all galaxies of Sample 1 to a constant scale before decomposing them (see \S~\ref{decomp}).  These galaxies are included in the SDSS Early Data Release.

\begin{figure}
\begin{center}
\scalebox{0.8}{\rotatebox{90}{\plotone{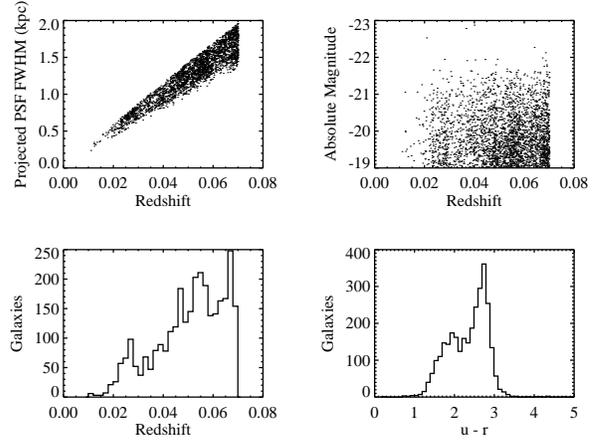}}}
\caption{The distributions of some physical quantities for Sample 2. \label{fig03}}
\end{center}
\end{figure}

We next investigate the shapelet method on the $r$-band images of a volume-limited sample of 3037 nearby galaxies (hereafter Sample 2) included in the SDSS DR1 \citep{aba03}.  These galaxies were chosen because they have projected PSF widths of less than our desired resolution of 2.0 kpc, allowing use to smooth them to this scale.  They have redshifts $z<0.07$ and absolute magnitudes $R<-19$.  We made the redshift cut at $z=0.07$ because projected PSF widths become larger than 2.0 kpc for redshifts greater than this, and we chose galaxies of absolute magnitudes $R<-19$ to allow a uniform distribution of absolute magnitude with redshift.  Figure \ref{fig03} displays the $r$-band PSF, redshift, and $r$-band absolute magnitude distributions for this sample.

\section{DECOMPOSITION METHOD USING SDSS DATA}

\label{decomp}

\begin{figure}
\begin{center}
\scalebox{1.5}{\rotatebox{90}{\plottwo{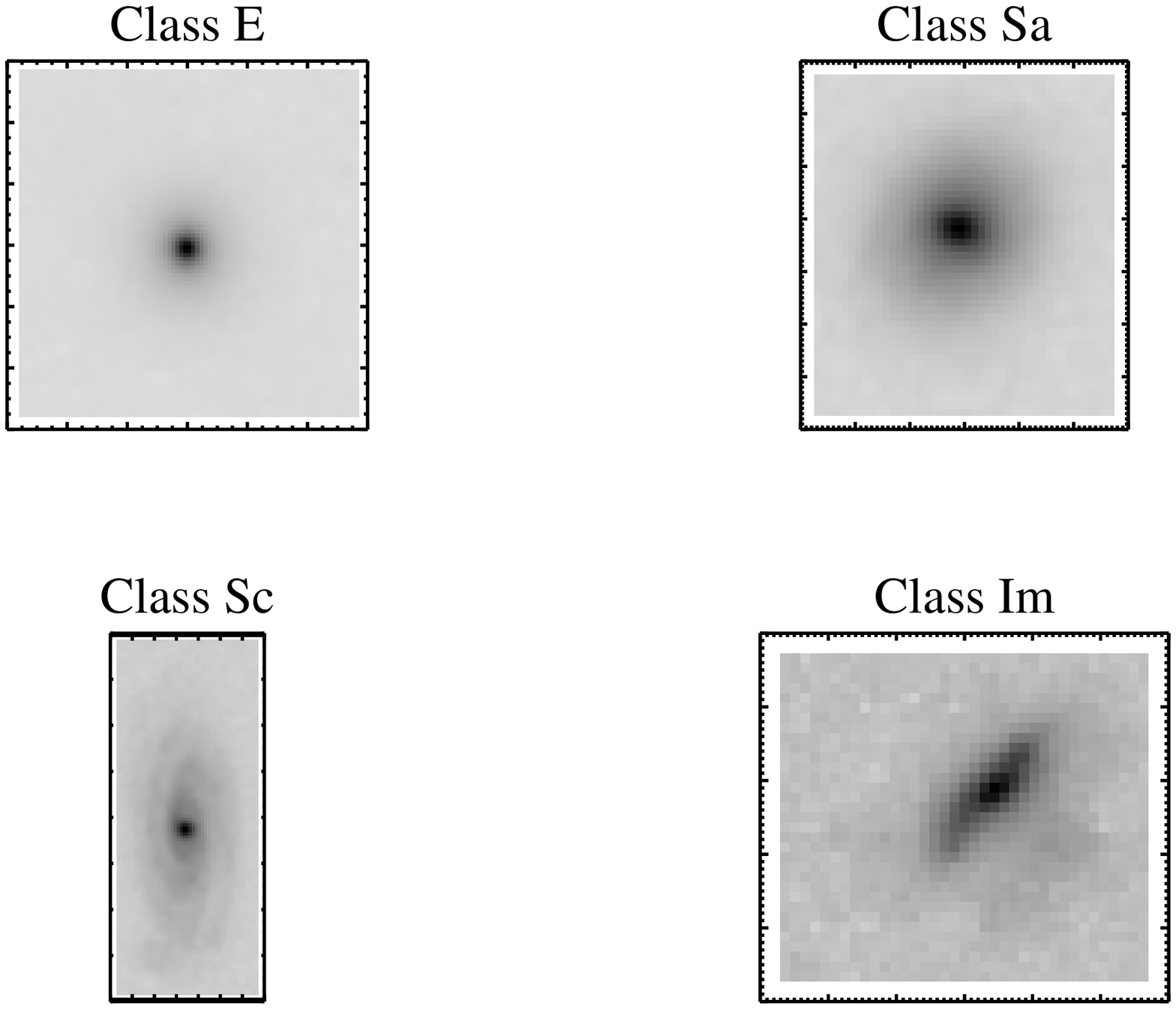}{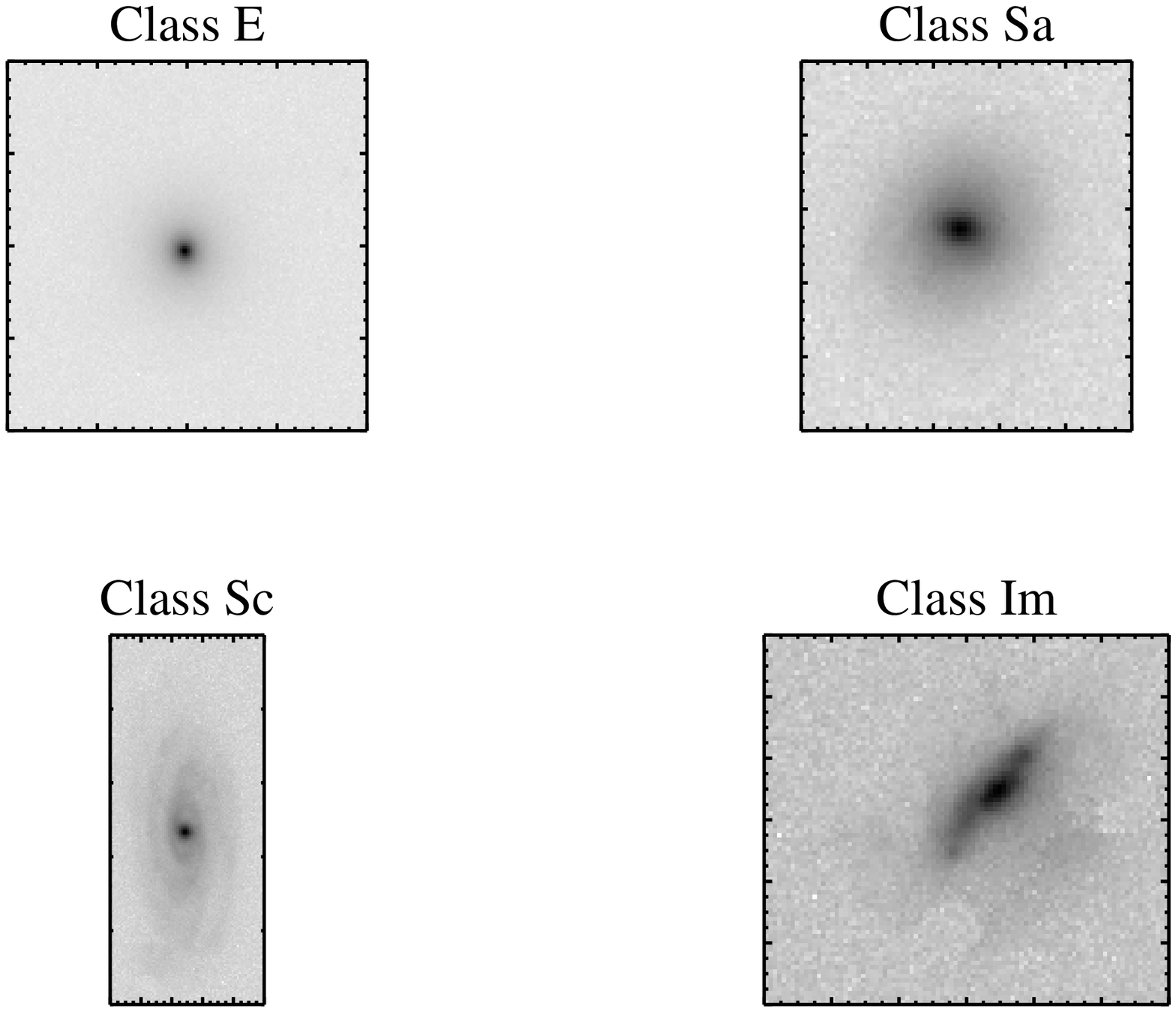}}}
\caption{Selected galaxies from Sample 1, shown before artificial redshifting and smoothing (top) and after (bottom). \label{fig04}}
\end{center}
\end{figure}

We describe here how we decompose galaxies using SDSS data and the shapelet formalism.  First, we input the atlas image $h({\bf x})$ for a given color band ({\it u,g,r,i,} or {\it z}), as well as the adaptively weighted moments in the horizontal and vertical directions, $I_{xx}$ and $I_{yy}$ \citep{fisc00}.  The atlas images are cutouts that include all significant pixels around each detected galaxy \citep{sto02}.  The atlas images of several galaxies classified according to Hubble type are shown in Figure \ref{fig04}.  We use the cataloged SDSS data to find the image center ${\bf x}_C$ and the position angle $\theta_{pos}$ from the horizontal.  We rotate the image an angular distance $\theta_{pos}$ so that each has its major axis along the horizontal.  We compute the characteristic scale $\gamma$ of our analyzing shapelets, given by
\begin{equation}
\gamma= \frac{\sqrt{I_{xx}+I_{yy}}}{2}.
\label{eq15}
\end{equation}
If the scale $\gamma$ (in pixels) was found to be less than $\beta$ pixels, where $\beta$ is the scale of the PSF, then we set $\gamma = \beta$.  If necessary we pad the image with blank sky out to a radius of $10 \gamma$ to assure the orthogonality of the basis functions.  We add artificial noise to the padded regions using the recorded value for $\sigma_{sky}$, and then subtract the mean of the image.

To assure uniformity within the sample, we artificially redshift all galaxies to $z=0.07$ and smooth them to a resolution that would be seen with a PSF $g_0 ({\bf x})$ of standard physical width $\beta_0$.  Doing so keeps the physical scale of resolution for all galaxies in our sample constant, assuring that information that may contribute to classification (i.e., separation in shapelet space) is not affected by differences in resolution.  Artificial redshifting is accomplished by rebinning the galaxy image so that it occupies fewer pixels.  We neglect the changes in the ratio of the galaxy flux to sky flux, as well as the increase in noise (e.g., from the galaxy, sky, dark current, etc.), that would be seen at higher $z$.  Such changes are negligible because artificially redshifting galaxies out to redshift of only $z=0.07$ does not produce any noticeable changes in these quantities.  Also, we do not correct for the decrease in galaxy flux that results from artificially redshifting the galaxy, since we normalize the image so that its total flux is unity, making flux correction unnecessary.

Smoothing a galaxy by a PSF of width $\beta_0$ amounts to performing the convolution
\begin{equation}
h'({\bf x}) = \int f({\bf x-x'}) g_0({\bf x'}) d^2 x, 
\label{eq16}
\end{equation}
where $h'({\bf x})$ is the image after convolving with a PSF $g_0 ({\bf x})$ of standard width $\beta_0$, and $f({\bf x})$ is the galaxy.  Because we do not observe $f({\bf x})$ directly, but rather $h({\bf x}) = \int f({\bf x-x'}) g({\bf x}) d^2 x$, where $g({\bf x})$ is the actual PSF, we must develop a way to calculate $h'({\bf x})$ from $h({\bf x})$.  This may easily be done in shapelet space, where convolution is represented by multiplication \cite[see][]{ref03b}. If we approximate the actual PSF $g({\bf x})$ as a Gaussian of scale $\beta$, we can compute in $x$-space the image $h'({\bf x})$ smoothed to the standard scale $\beta_0$.  We do this by performing the convolution
\begin{equation}
h'({\bf x}) = \int h({\bf x-x'})g'({\bf x}) d^2 x,
\label{eq17}
\end{equation}
where $g'({\bf x})$ is a Gaussian of scale $\beta'=\sqrt{\beta_0^2 - \beta^2}$.  If one is using a non-Gaussian smoothing kernel (PSF) or wishes to compute the $h'_{\bf n}$ after decomposition then one must do the computation in shapelet space.  Although the PSFs from the SDSS data are not exactly Gaussian, the error in the coefficients $h'_{\bf n}$ introduced from approximating them as so is negligible for our purposes. The scale of the shapelets used in the decomposition is $\gamma'=\sqrt{\gamma^2 - \beta^2 + \beta_0^2}$, where $\gamma'$ is used to correct for the loss of resolution.  This assures that $\gamma' \ge \beta_0$.

\begin{figure}[t]
\begin{center}
\scalebox{0.8}{\rotatebox{90}{\plotone{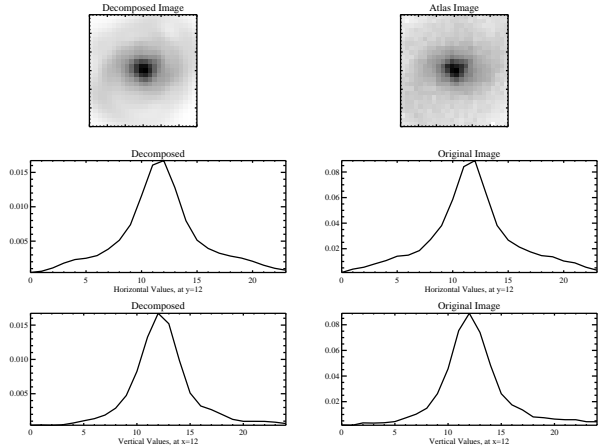}}}
\caption{A late type galaxy image (right), and the image reconstructed from the shapelet coefficients $f_{\bf n}$ (left), for $n_{max}=n_1+n_2=12$.  The galaxy image was decomposed according to the method outlined in \S~\ref{decomp}. Also shown are the horizontal (middle) and vertical (bottom) cross-sections of the images, taken at half the opposite axis.  The shapelet method accurately reconstructs the original image. \label{fig05}}
\end{center}
\end{figure}

Using Equations (\ref{eq06}) and (\ref{eq07}), and our values for $\gamma'$ and ${\bf x}_C$, we now calculate the shapelet coefficients by decomposing the galaxy image about its center, up to a maximum order $n_{max} = n_1+n_2$.  Because centering errors are introduced from the rotation, resizing, and smoothing of the image, and to assure the integrity of our decomposition, we compute the center of the image from Equation (\ref{eq09}) using these coefficients and redecompose the galaxy about this new center; we do this twice.  We then compute the flux of the decomposed image from the shapelet coefficients (Eq.[\ref{eq08}]) and divide the coefficients $h'_{\bf n}$ by the flux.  This normalizes the shapelet coefficients such that the flux of the image reconstructed from them is equal to unity.  Finally, we convert the Cartesian shapelet coefficients to the polar ones using Equation (\ref{eq14}).  Figure \ref{fig04} also shows the Hubble-classified galaxies after artificial redshifting and smoothing, and Figure \ref{fig05} shows the reconstruction of a galaxy from its shapelet coefficients.

In this analysis we choose the standard physical width $\beta_0$ to be 2.0 kpc.  A $\beta_0$ of 2.0 kpc allows us to use a large number of galaxies ($\sim 3000$) out to a redshift of $z=0.07$ (see Figure \ref{fig03}), resulting in a sample that is a good representation of galaxies that populate the nearby universe.  In addition, a smoothing scale of $\beta_0 = 2.0$ kpc allows the large-scale structures of galaxies to be sufficiently well resolved for classification.  Small-scale structures do not result in appreciable differences in shapelet coefficients within the context of morphological classification, and losing this information does not affect our classification method.  We choose $n_{max} = 12$, assuring a sufficiently accurate reconstruction for even those galaxies with the most complicated morphologies, while keeping the computation time reasonable.

\section{PRINCIPAL COMPONENT ANALYSIS}

\subsection{The Most Powerful Shapelet States}

\label{top10}

We decompose the galaxies of Sample 2 according to the procedure outlined in \S~\ref{decomp}.  In order to have an understanding of which shapelets consistently contain the most power, and thus give the best description of the most common aspects of the morphologies of nearby galaxies ($z < 0.07$), we have calculated the top ten most powerful shapelet states for all the galaxies in Sample 2.  These states are shown in Table \ref{tab01} for both the Cartesian and Polar shapelets, and several of them can be seen in Figures \ref{fig01} and \ref{fig02}.  As expected, the symmetric states are the most powerful, with the first order asymmetric state the only asymmetric one that appears among the top ten most powerful states.  In the case of the Cartesian states, we see that the states with more horizontal structure ($n_1 > n_2$) are more powerful than those with more vertical structure ($n_2 > n_1$).  This is the result of the way in which we reduce the images before decomposing them, as we rotate the image such that its major axis is oriented along the horizontal.  The fact that the horizontal states are more powerful than the vertical is merely a representation of the fact that there is more flux and structure along a galaxy's major axis than its minor.  Also, the state $|0, 0 \rangle$ (a Gaussian) is the same for both the polar and Cartesian representations, and contains a little less than one-third of the total power for both.

\begin{table}[t]
\begin{center}
\begin{scriptsize}
\begin{center}
\caption{Top Ten Most Powerful Shapelet States.\label{tab01}}
\end{center}
\begin{tabular}{ccccccc}
\tableline \tableline
 & \multicolumn{3}{c}{Cartesian\tablenotemark{a}} & \multicolumn{3}{c}{Polar\tablenotemark{b}} \\
Rank & State & $\langle |f_{\bf n}| / f_{tot} \rangle$\tablenotemark{c} 
& $\sigma_{\bf n}$\tablenotemark{d} & State & 
$\langle |f_{\bf n}| / f_{tot} \rangle$ & $\sigma_{\bf n}$ \\
\tableline
1  & $|0,0 \rangle$ & 0.293 & 0.076 & $|0,0 \rangle$ & 0.309 & 0.089 \\
2  & $|4,0 \rangle$ & 0.047 & 0.012 & $|4,0 \rangle$ & 0.057 & 0.020 \\
3  & $|2,0 \rangle$ & 0.059 & 0.021 & $|2,0 \rangle$ & 0.050 & 0.029 \\
4  & $|0,4 \rangle$ & 0.026 & 0.011 & $|2,-2 \rangle$& 0.029 & 0.018 \\
5  & $|6,0 \rangle$ & 0.022 & 0.010 & $|2,2 \rangle$ & 0.029 & 0.018 \\
6  & $|2,2 \rangle$ & 0.021 & 0.010 & $|8,0 \rangle$ & 0.021 & 0.015 \\
7  & $|0,2 \rangle$ & 0.030 & 0.020 & $|6,0 \rangle$ & 0.023 & 0.017 \\
8  & $|8,0 \rangle$ & 0.018 & 0.009 & $|1,-1 \rangle$& 0.017 & 0.012 \\
9  & $|0,1 \rangle$ & 0.016 & 0.015 & $|10,0 \rangle$& 0.016 & 0.012 \\
10 & $|1,0 \rangle$ & 0.014 & 0.012 & $|1,1 \rangle$ & 0.017 & 0.012 \\
\tableline
\end{tabular}
\tablenotetext{a}{Cartesian states are labeled as $|n_1, n_2 \rangle$.}
\tablenotetext{b}{Polar states are labeled as $|n, m \rangle$.}
\tablenotetext{c}{Mean ratio of the coefficient power to the total coefficient power}
\tablenotetext{d}{Standard Deviation in the coefficient power ratio}
\end{scriptsize}
\end{center}
\end{table}

For the polar shapelets, it is obvious that the states with azimuthal symmetry ($m=0$) are the most powerful, and again this is not surprising.  This reflects the fact that the galaxies of Sample 2 exhibit more radial structure than angular structure.  The only $m \neq 0$ states are $|2, \pm 2 \rangle$ and $|1, \pm 1 \rangle$. The former state picks up the ellipticity of a galaxy, and the latter state is just the first order asymmetry state.

\subsection{Principal Component Analysis}

\label{pca}

Performing the shapelet decomposition for $n_{max} = 12$ results in 91 coefficients for each galaxy; 91 dimensions in the shapelet coefficient space.  Such a high dimensionality presents a significant barrier to understanding the nature of the galaxy distributions in shapelet space, as well as performing computations in it (e.g., density estimations).  In order to reduce the dimensionality of the space we perform a principal components analysis.  The principal component transform (also called a Karhunen-Loeve transform) of a set of $p$-dimensional data is a linear transformation that computes the $q$ principal components, for $q \leq p$.  The principal components of a data set are very useful for reducing the dimensionality of the data.  This is because the $q$ principal components provide the best rank $q$ linear approximation to the data, and provide an orthogonal basis that maximizes the variance along the components; i.e., the first principal component is in the direction of highest variance in the data, the second principal component is in the direction of highest variance subject to being orthogonal to the first principal component, etc. \citep{hast01}.  In the context of the work done here, the principal components are linear combinations of shapelet states, and can be constructed using Equation (\ref{eq06}) to form a 2-dimensional image.  We denote the $j^{th}$ principal component by $p_j ({\bf x})$ and its corresponding coefficient by $a_j$.

We apply the Karhunen-Loeve transform on the coefficients of the Cartesian states for the galaxies of Sample 2 and find that 98.7~\% of the variance is contained within the first three principal components, with 96.7~\% within the first principal component.  This implies that the information in the shapelet decomposition is incredibly redundant, and allows us to decrease the dimensionality of the space to something that is much easier to work with.  We found that there was no noticeable difference between applying the transform to the Cartesian or polar coefficients.  Although the polar coefficients are complex, the majority of the shape information is contained within the real part of the polar coefficients.  This means that performing the transform on the real part of the polar coefficients gives nearly the same result as applying it on the Cartesian coefficients.  This is not surprising, as the conversion from Cartesian to polar coefficients is itself a linear transform.  We choose to apply the principal component analysis to the Cartesian coefficients for convenience and to ensure that we do not lose any information, however insignificant, as would happen if we were to use the polar coefficients and ignore the imaginary part.

\begin{figure}
\begin{center}
\scalebox{1.0}{\rotatebox{90}{\plotone{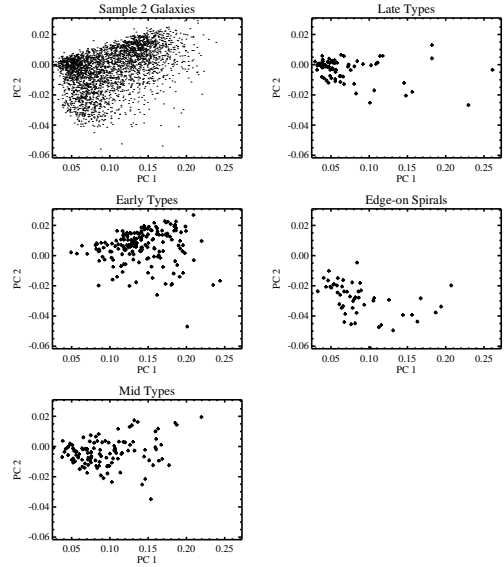}}}
\caption{Locations of the galaxies of Sample 2 (top left plot) and Sample 1 in the 2-dimensional slice spanned by $p_1$ and $p_2$.  Using the reference galaxies from Sample 1, we are able to see that the different Hubble types are well separated in this plane. \label{fig06}}
\end{center}
\end{figure}

\begin{figure}
\begin{center}
\scalebox{0.8}{\rotatebox{90}{\plotone{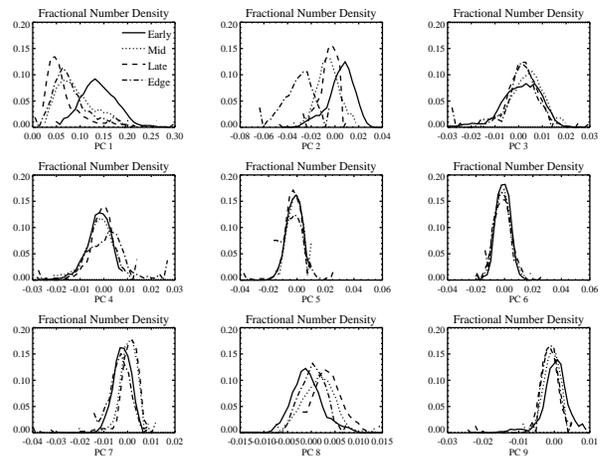}}}
\caption{Projections of the galaxies of Sample 1 along the first nine principal components. From these plots, one can see that the Hubble types are well seperated along a few of the $p_j$. \label{fig07}}
\end{center}
\end{figure}

\begin{figure}
\begin{center}
\scalebox{0.8}{\rotatebox{90}{\plotone{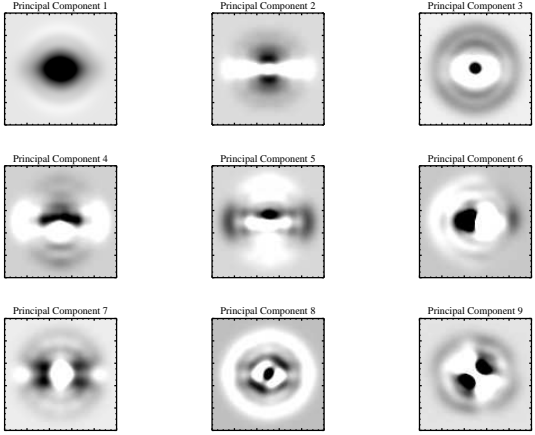}}}
\caption{Images of the first nine principal components, constructed from their shapelet coefficients. \label{fig08}}
\end{center}
\end{figure}

We decomposed the galaxies of Sample 1 in the same manner as for Sample 2.  We keep those galaxies of Sample 1 with absolute $r$-band magnitude $R > -19$, since we are only using Sample 1 for reference and these galaxies are not included in the principal component transform or the density estimation (see \S~\ref{mixclass}). We divide the galaxies of Sample 1 into four types---early, mid, late, and edge-on.  The early types include those Hubble-classified by \citep{nak03} as E--Sa, the middle types include Sab--Sbc, and the late types include Sc--Im (note that Im galaxies are classified as `irregular').  The edge-on class contains those mid and late type galaxies with axis ratios $b/a < 0.4$, where $b$ and $a$ are the minor and major axis, respectively.  Only a few of the $p_j$'s exhibited structure or separation of Hubble type for the galaxies of Sample 1.  We use the first nine principal components, which includes those that showed Hubble type separation.  This allows us to work within a 9-dimensional space spanned by $\{ p_1, p_2, p_3, p_4, p_5, p_6, p_7, p_8, p_9 \}$, greatly simplifying our analysis.  It should be noted that not all of a galaxy's morphological information is contained within these principal components, but rather that a sufficient amount of information pertaining to the important {\em differences} in galaxy morphology is contained in them.  It is only the information regarding differences in morphology that is useful for classification, which is why we use these nine principal components.  Principal components after $p_9$ became increasing irregular and likely only serve to include small-scale information unique to each galaxy.  Locations of the galaxies in the 2-dimensional slice spanned by the first two principal components are shown in Figure \ref{fig06}, the projections of the galaxies along the $p_j$ are shown in Figure \ref{fig07}, and the first nine principal components, $p_j ({\bf x})$, are shown in Figure \ref{fig08}.

\subsection{Description of the Principal Components}

\label{pcdesc}

Comparing the projections along the principal components with the images of the $p_j$ allows us to better understand why the galaxies separate in this space in such a manner.  The first Principal Component, $p_1 ({\bf x})$, accounts for the vast majority of the variance ($\approx~97~\%$).  It is slightly elliptical and has mirror symmetry.  We have fit the profile of $p_1$ and find that it is exponential, except near the edge.  Comparison with the other principal components reveals that the coefficients of $p_1 ({\bf x})$, $a_1$, are typically several times larger in magnitude than those for the other $p_j ({\bf x})$.  This implies that $p_1 ({\bf x})$ forms what we may consider to be the basic galaxy shape, while the other $p_j$ are subsequent modifications to it.  This is reinforced by the fact that the coefficients and the flux of $p_1$ are always positive, ensuring that galaxies do not contain `holes' of negative flux.  The late types typically have the least power in this component relative to the other principal components, while the early types typically have the most.  This implies that `more' of the morphology of early types is contained within $p_1 ({\bf x})$ than for late types, i.e., a greater fraction of the total power in the coefficients for early types is contained within this principal component.  Because early types typically have the simplest morphologies, this result is not surprising, since $p_1$ certainly contains minimal structure.  Since a greater fraction of the shapelet power is contained within $p_1$ for early types (small $|a_j| / a_1$), fewer corrections are needed to this shape for early types than for later types.  In fact, most of the corrections to $p_1$ for early types appear to be for the purpose of changing the shape of the radial light profile.

\begin{figure}
\begin{center}
\scalebox{0.8}{\rotatebox{90}{\plotone{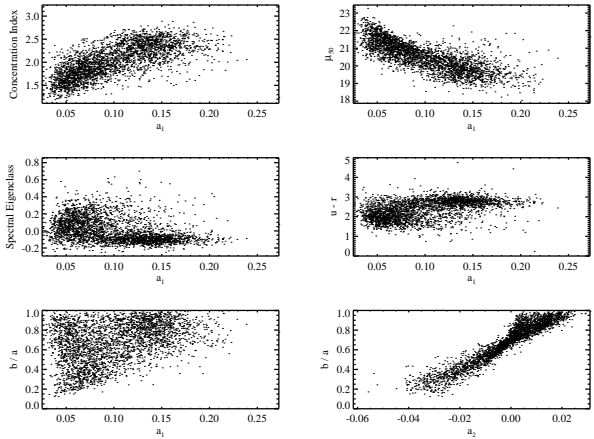}}}
\caption{Plots of the coefficients for the principal components as a function of several physical parameters.  The correlations between $p_1$ and $\mu_{50}$ and $p_2$ and $b / a$ are obvious. \label{fig09}}
\end{center}
\end{figure}

In Figure \ref{fig09} we show the dependence of several physical and morphological parameters as a function of the coefficients of $p_1$.   One can see that a galaxy's concentration index, a common morphological measurement, correlates well with its projection along $p_1$.  We define the concentration index $C$ as
\begin{equation}
C = 5.0 \log_{10} \left ( \frac{r_{90}}{r_{50}} \right )
\label{eq18}
\end{equation}
where $r_{90}$ and $r_{50}$ are the radii where the Petrosian ratio $\eta$ is equal to 0.1 and 0.5, respectively.  The Petrosian ratio is the ratio of the local surface brightness at some radius, as averaged over an annulus, to the average surface brightness within that radius.  More concentrated galaxies (such as early types) will have higher values of $C$.  In addition, the $r$-band half-light surface brightness $\mu_{50}$ also appears to be correlated with $a_1$.  We define $\mu_{50}$ to be the average surface brightness within the Petrosian half-light radius ($\eta = 0.5$).  The units are magnitudes per arcsec$^2$.  Although there is a significant spread in $\mu_{50}$ as a function of $a_1$, a broad linear is apparent; surface brightness increases (decreases in magnitudes) as $a_1$ increases.  We interpret this as meaning that galaxies with the simplest morphologies, at least as seen when resolved on the same physical scale, tend to have the highest surface brightness.  These galaxies also tend to be the most concentrated, and this relationship between concentration and surface brightness is in accordance with previous studies \citep{blan03}.

We also show the relationship between $a_1$ and some of a galaxies spectral properties, by plotting $u - r$ color and spectral eigenclass as functions of $a_1$.  The spectral eigenclass is the negative of the mixing angle, $\phi_{KL}$, between the first and second spectral eigencoefficients, where the spectral eigencoefficients, $e_i$, are found from a principal component analysis of 170,000 SDSS spectra \citep{yip03}.  The mixing angle is $\phi_{KL} = \tan^{-1} (e_2 / e_1)$, and the spectra cover the range 3450--8350 \AA.  Red (elliptical) galaxies typically have negative spectral eigenclass values, and blue (spiral) galaxies typically have positive values. The first eigenspectrum is the mean of all galaxy spectra, with a continuum similar to a Sb-type.  The second eigenspectrum is negative below $\sim 5200$~\AA, positive at longer wavelengths, and contains the majority of the stellar absorption information.  The third eigenspectrum (used in \S~\ref{mix_descript}) is positive below $\sim 6000$~\AA, negative at longer wavelengths, and is dominated by line emission.  One can see that less concentrated galaxies with more complex morphologies (i.e., smaller $a_1$) have a broader range of spectral properties and are often bluer than those galaxies with simpler, more concentrated shapes, but show a narrower range of $a_1$ than the red galaxies.  The coefficient of $p_1$ does not appear to have a strong dependence on axis ratio other than the fact that the distribution in the projections along $p_1$ narrows with increasing ellipticity.  Relationships between these physical and morphological parameters were also examined for the other $p_j$, but the structure was not as well defined.  We do not discuss them because the $a_j$ are often correlated with one another, making the discussion redundant.

The second principal component, $p_2 ({\bf x})$, contains a significant amount of the ellipticity information of a galaxy.  As can be seen in Figure \ref{fig09}, the ellipticity increases with decreasing coefficient of $p_2$ in a well-defined, linear manner.   The edge-ons separate well along this principal component, and have the most powerful coefficients of $p_2 ({\bf x})$; in fact, $p_2 ({\bf x})$ resembles very much what we would expect an edge-on spiral galaxy to look like, with the exception of negative flux and the `peaks' seen in the vertical direction.  These `peaks' serve to narrow the galaxy in the vertical direction, since the positive values for the pixels in this region are multiplied by the $a_2$, which are negative for edge-ons, and added to the positive values for the pixels from $a_1 p_1({\bf x})$.  The early types also separate very nicely along this component.  The early types are found to mostly have positive values for $a_2$. This may seem odd at first, however, inspection of $p_1 ({\bf x})$ and $p_2 ({\bf x})$ reveal why this is the case.  The first principal component is slightly elliptical.  Early type galaxies of ellipticity the same as that of $p_1$ should not require any modification from $p_2$, and thus would have projections along $p_2$ of nearly zero magnitude.  However, if the early type has ellipticity less than that of $p_1$, it will need the additional contribution from $p_2$ to elongate it in the vertical direction and narrow it in the horizontal direction.  In the case where the ellipticity is less than that of $p_1$, the coefficient is negative, and the reverse of $p_2 ({\bf x})$ is used to modify $p_1$.  This results in the two symmetric `peaks' in the vertical direction becoming `valleys', and thus would widen the contribution from $p_1$, making the galaxy more elliptical.  The coefficients of $p_2$ for the middle and late types are distributed preferentially negative, but with less magnitude than the $a_2$ for the edge-ons. This implies that, on the average, the late and middle types are slightly more elliptical than $p_1$.

\begin{figure}[t]
\begin{center}
\scalebox{0.8}{\rotatebox{90}{\plotone{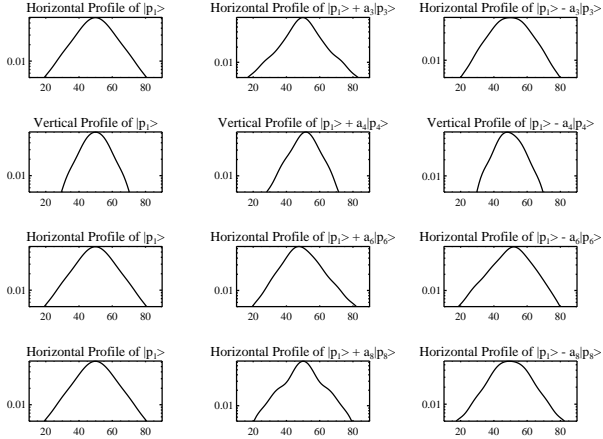}}}
\caption{Light profile of $p_1 ({\bf x})$ before (left column) and after addition (middle column) and subtraction (right column) of selected principal components.  Values of $|a_j|/a_1$ used were chosen to be large enough to emphasize the contributions from the $p_j$, but still characteristic of real galaxies. The vertical axis is shown logarithmically. \label{fig10}}
\end{center}
\end{figure}

The other principal components, with the exception of $p_7$ and $p_8$, do not separate the Hubble types as well as the first two, nor do the distributions appear to have as much structure.  The third and eighth principal components have light profiles that are more centrally concentrated than the previous two, as well as alternating annuli of negative and positive flux around the central peak. Late types typically have positive values for $a_8$, while early types have negative values.  This modification for late types results in a narrowing of the radial profile near the center and a widening farther out.  For early types, $p_8$ flattens the light profile near the center and makes it steeper farther out.  The effect of $p_3$ is similar, with the exception that it does not contain as much radial structure as $p_8$.  The contributions of these two principal components help to define the shape of a galaxy's light profile, and in the case of late types may contain morphological information regarding the separation of the bulge and disk components.  The light profile of $p_1 ({\bf x})$, as well as the result after the addition of $p_3 ({\bf x})$ and $p_8 ({\bf x})$ for both positive and negative $a_j$ are shown in Figure \ref{fig10}.

The fourth, fifth and sixth principal components contain information regarding asymmetry about the major ($p_6$) and minor axis ($p_4$ and $p_5$).  The fourth and fifth principal components appear to have odd symmetry with each other about the major axis, i.e., $p_4 (- {\bf x}) = - p_5 ({\bf x})$, and also contain further information regarding ellipticity.  In spite of the similarities in the morphologies of $p_4$ and $p_5$, their coefficients do not appear to be correlated.  The coefficients of $p_4, p_5,$ and $p_6$ may be thought of as giving a weighted asymmetry measurement.  This is because the $a_j$ are given by the usual inner product as $a_j = \langle p_j ({\bf x}) | f ({\bf x}) \rangle$, which is equal to the sum of the values of the pixels of our galaxy image weighted by the value of the corresponding pixel of the $p_j$ image.  In this case, preference is given to asymmetry close to the center, where the principal components have the highest flux.  Because of this, these principal components do not do a good job of picking out galaxies where the asymmetry is far from the center and of lower surface brightness.  The coefficients of the galaxies for these $p_j$ appear to be symmetrically distributed around zero regardless of Hubble type, as would be expected. The light profiles after the contributions from these principal components are also shown in Figure \ref{fig10}.

Hubble types appear to have a small separation in $p_7$, with the early and edge-on galaxies having negative $a_7$ and the middle and late type galaxies having positive $a_7$.  The morphological contribution from $p_7$ appears to be small; for galaxies with positive $a_7$ it may serve to make them more `boxy' along their major axis, as well as correcting the light profile to make it more exponential far from the center, where $p_1$ diverges from an exponential profile.  For galaxies with negative $a_7$, it may contribute to the creation of a low surface brightness disk viewed edge-on, such as would be seen in an S0 or Sa galaxy.

The ninth principal component does not significantly separate Hubble types.  The contribution from $p_9$ is small, and the values of $a_9$ are symmetric about zero.  The morphological contribution from $p_9$ is to make the central isophotes more `boxy' and tilted at a $\pm 45$ degree angle from the major axis, where the tilt is dependent on the sign of $a_9$.

\section{MIXTURE OF GAUSSIANS MODEL AND CLASSIFICATION}

\subsection{The Statistical Model}

\label{mixmod}

Now that we have a 9-dimensional representation of the data, we wish to have a mechanism that identifies clusters in the data set.  Each of these clusters then corresponds to a particular morphological class.  We use a mixture of Gaussians model to estimate the galaxy density in shapelet space, and to form the basis of our classification system.  To implement the mixture model method, we use the publicly available {\bf fastmix} software \citep{moo99}, which employs multi-resolutional KD-trees to overcome many computational issues associated with mixture models.  Here, we outline the important aspects of the method found in \citet{con00}.  The basic idea behind the mixture of Gaussians model is to regard each $d$-dimensional data vector $X_i$ as being drawn from a probability density $F$.  Because the integral of $F$ over the region occupied by the $X_i$ is unity, i.e., the probability of drawing each $X_i$ over the entire range of the data is one, the probability density $F$ is the normalized number density for the data.  The distribution is modeled as
\begin{equation}
F(x; \theta_k) = \sum_{j=1}^k \tau_j \Phi(x; \mu_j, \Sigma_j),
\label{eq19}
\end{equation}
where $\Phi$ denotes a $d$-dimensional Gaussian of mean $\mu$ and covariance $\Sigma$.  Each $\Phi$ is given by
\begin{equation}
\Phi (x; \mu, \Sigma) = \frac{1}{(2 \pi)^{d/2} | \Sigma |^{1/2}} \exp \left [ - \frac{1}{2} (x - \mu)^{T}
\Sigma^{-1} (x - \mu) \right ].
\label{eq20}
\end{equation}
The parameters to be estimated from the data set are the number of Gaussians $k$ and $\theta_k=(\tau, \mu, \Sigma)$, where $\tau=(\tau_0, \ldots, \tau_k)$, $\mu=(\mu_0, \ldots, \mu_k)$, and $\Sigma=(\Sigma_0, \ldots, \Sigma_k)$.  The values of $\tau$ are limited such that $\tau_j \geq 0$ for all $j$ and $\sum_{j=0}^{k} \tau_j = 1$.

To estimate the $\theta_k$ the method of maximum likelihood is used.  Assuming that $k$ is fixed and known, this amounts to finding the $\theta_k$ that maximizes the likelihood function.  The likelihood function is defined as the probability of the observed data $X_i$ as a function of the unknown parameters $\theta_k$.  In order to find the $\theta_k$ that maximizes the likelihood function, {\bf fastmix} employs the expectation maximization (EM) algorithm.  To choose $k$, we opt for using the Bayesian Information Criterion (BIC, \citet{sch79}).  The BIC criteria corresponds to approximately maximizing the posterior probability of the $k^{th}$ model (in this case $F(x; \theta_k)$), and gives preference to simpler models \citep{hast01}, e.g., a mixture of Gaussians model with fewer Gaussians.

\subsection{Classification Using the Mixture Model}

\label{mixclass}

The mixture of Gaussians model provides a natural basis for the use of morphological galaxy classification.  We expect that a galaxy's morphology is the result of its evolution, and galaxies of similar morphology, i.e., in the same morphological class, have undergone similar events in their evolution to bring about their morphology (e.g., mergers, collisions, spiral density waves, etc.).  We also expect galaxies to have histories which are the result of physical processes unique to each galaxy, resulting in small variations about the `mean' morphology of a class.  We assume that these variations may be taken as random when looking over a large enough sample of galaxies.  This results in the morphologies of galaxies in the same class being normally distributed about the average morphology.  If the morphology of galaxies is normally distributed about the mean shape, then the shapelet coefficients $f_{\bf n}$ are also normally distributed about the mean coefficient, $\langle f_{\bf n} \rangle$.  A mixture of Gaussians model then provides the most natural basis for classification, where galaxies of the same Gaussian are of the same class.  Galaxies may be assigned a probability for each Gaussian, allowing one to give a measure of how likely it is that the galaxy is in each class.  This continuous, quantitative description of classification is an advantage over the Hubble scheme, as it allows significant flexibility.  Rather than using discrete classes which do not allow for ambiguities, we are able to give a measure of exactly how confident we are that a given galaxy is in a given class, and ambiguous galaxies are not `pigeon-holed' into classes where they do not fit.  In addition, it should be noted that because we are interested in classification, rather than perfectly fitting the dataset, the BIC criteria allows for the most optimal fit while keeping the number of Gaussians (classes) reasonable. 

We use the 9-dimensional space spanned by the principal components $\{ p_j, j = 1, \ldots, 9 \}$ as the morphological space in which to work.  To estimate the density distribution we used the coefficients of the principal components for Sample 2 and fed them into {\bf fastmix}.  Because most of the $p_j$ contain axis ratio information, and what we see in reality is a 2-dimensional projection of the galaxy, where the galaxy is tilted by an angle from our line of sight, we do not expect the observed galaxy morphologies to be perfectly normally distributed.  This tilt distorts the galaxy morphologies, resulting in the shapelet coefficients diverging from a normal distribution.  This results in the need for extra Gaussians to fit the density, which may create spurious class where the only difference is that of axis ratio.  Fortunately, it appears that this effect is small and differences in axis ratio do not distort the shapelet coefficients enough to require extra Gaussians except for the case of spirals viewed edge-on or nearly edge-on.

\begin{figure}
\begin{center}
\scalebox{0.8}{\rotatebox{90}{\plotone{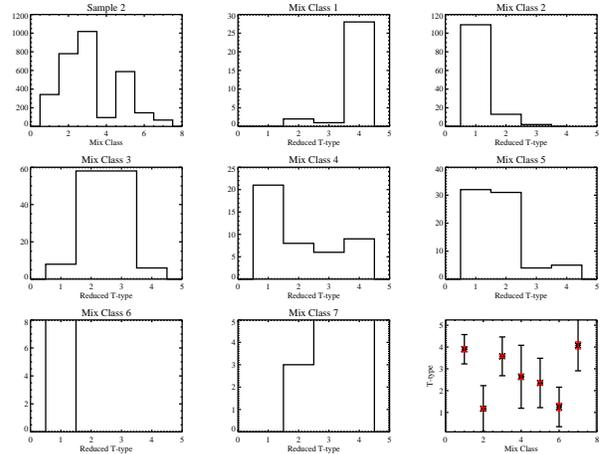}}}
\caption{Plots showing the results of the mixture model classification.  The first histogram (top left) shows the distribution of mixture classes for the galaxies of Sample 2.  The ``Reduced T-type'' seen in the following histograms are based on the Hubble classifications of Sample 1 and stands for the following: 1=Early, 2=Middle, 3=Late, 4=Edge-on.  Note that the $x$-axis is different for the first histogram because it shows the distribution of mixture classes, whereas the others show the distribution of broad Hubble type in each $M_k$.  The T-type values of the last plot are also from the Sample 1 data, and are as follows: 0=E, 1=S0, 2=Sa, 3=Sb, 4=Sc, 5=Sdm, 6=Im. As can be seen, the Hubble types map very well to several of the mixture classes.  The larger error bars seen in the bottom plots show the standard deviation in the distributions, and the shorter ones show the error in the mean.\label{fig11}}
\end{center}
\end{figure}

\begin{figure}
\begin{center}
\scalebox{0.8}{\rotatebox{90}{\plotone{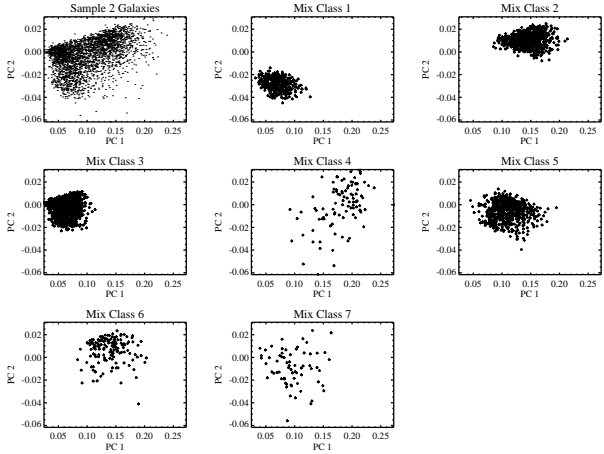}}}
\caption{Locations of the galaxies of Sample 2 in the 2-dimensional slice spanned by $p_1$ and $p_2$. \label{fig12}}
\end{center}
\end{figure}

\begin{figure}
\begin{center}
\scalebox{0.8}{\rotatebox{90}{\plotone{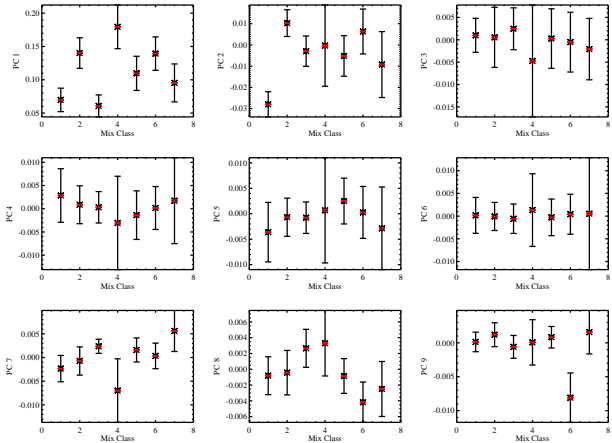}}}
\caption{Projections along the principal components for the galaxies of Sample 2, classified by mixture model class $M_k$. The larger error bars show the standard deviation in the distributions, and the shorter ones show the error in the mean. \label{fig13}}
\end{center}
\end{figure}

After applying the mixture model estimation, we find that the 9-dimensional density distribution is well fit with seven Gaussians.  Using this distribution, we are able to classify galaxies by assigning galaxies to the class (Gaussian) for which they have the highest probability, as well as assigning an associated probability of being in each class.  We compare the mixture class with Hubble type in Figure \ref{fig11}. In Figure \ref{fig12} we show the locations of the galaxies for each mixture class in the 2-dimensional slice spanned by $p_1$ and $p_2$, and in Figure \ref{fig13} we show the mean projection of the galaxies along the principal componetns. Analysis of the histograms in Figure \ref{fig11} reveals that the mixture model classifications do an excellent job of separating the galaxies based on their morphology, and we can certainly see from Figures \ref{fig12} and \ref{fig13} that the projections along the principal components for the largest mixture model classes correspond well with the Hubble types in Figures \ref{fig06} and \ref{fig07}.  This is encouraging, as it means that galaxies with commonly accepted morphological differences, i.e., of different Hubble type, form distinct classes in shapelet space.  The classes are found from an analytical model that does not admit any morphological assumptions (i.e., is model-independent), with the exception that the density distribution of galaxies in shapelet space can be well represented by a mixture of Gaussians.

\begin{figure}
\begin{center}
\scalebox{0.8}{\rotatebox{90}{\plotone{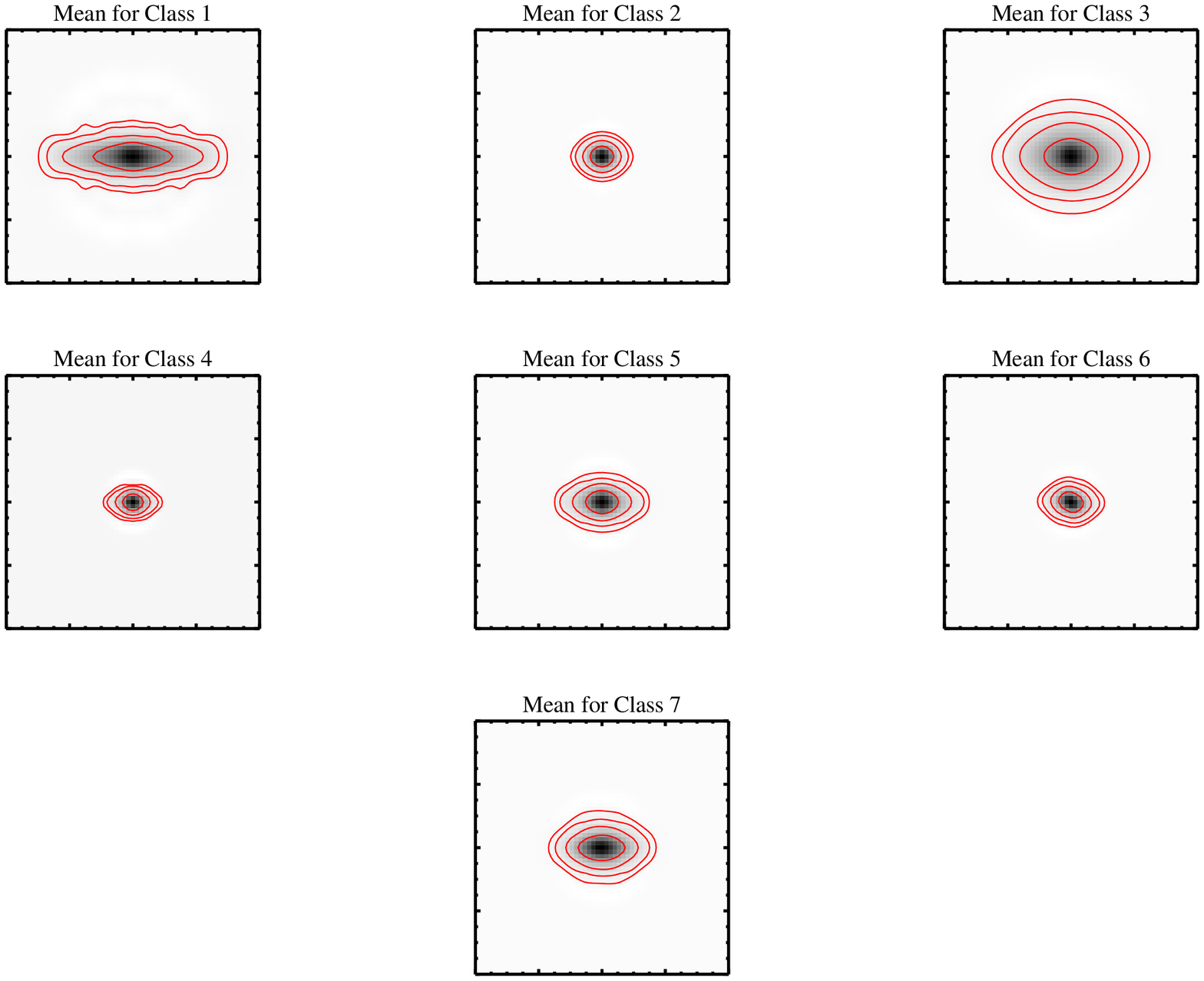}}}
\caption{The mean morphologies of the mixture model classes, reconstructed from the mean shapelet coefficients for each respective class.  The contours denote where the image drops to $e^{-j}$ of its maximum value, where $j$ is the contour number from the center.  For example, the first contour from the center labels where the flux has dropped to $e^{-1}$ of its maximum value. \label{fig14}}
\end{center}
\end{figure}

\begin{figure}
\begin{center}
\scalebox{1.7}{\rotatebox{90}{\plottwo{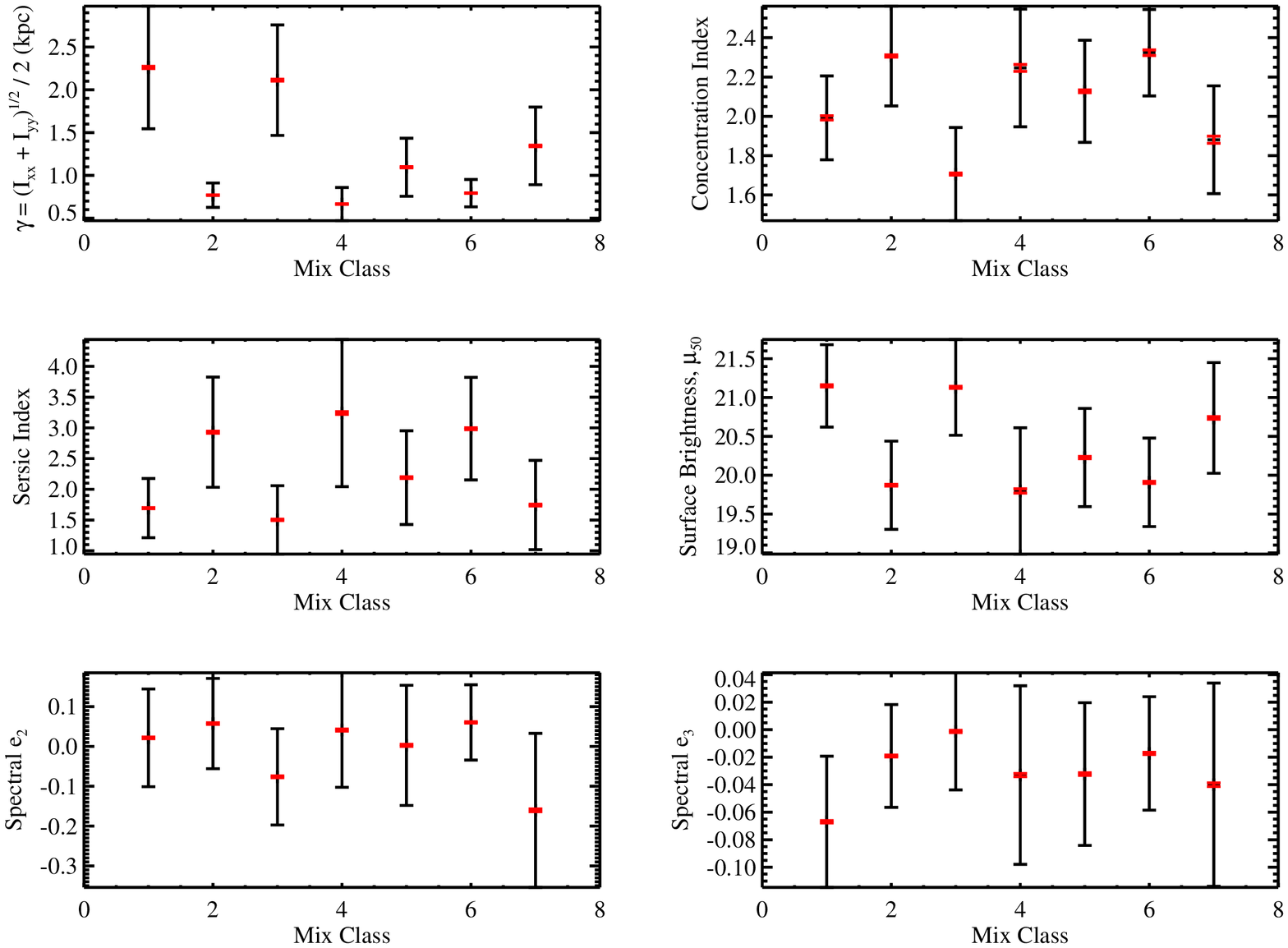}{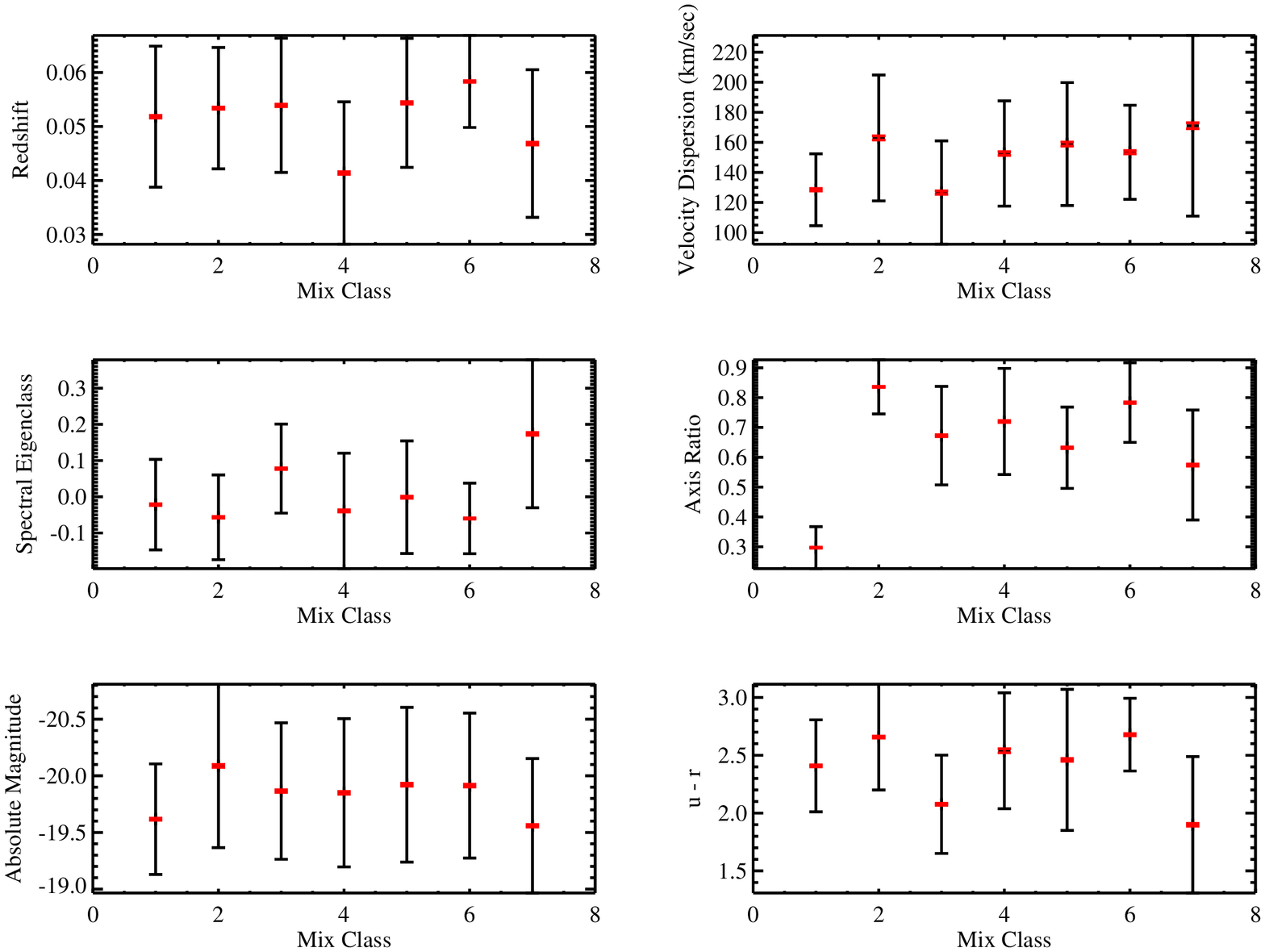}}}
\caption{The average of several physical properties for each Mixture class. The larger error bars show the standard deviation in the distributions, and the shorter ones show the error in the mean.  S\'{e}rsic Indices are from \citet{blan03}. \label{fig15}}
\end{center}
\end{figure}

\begin{figure}
\begin{center}
\scalebox{1.5}{\rotatebox{90}{\plottwo{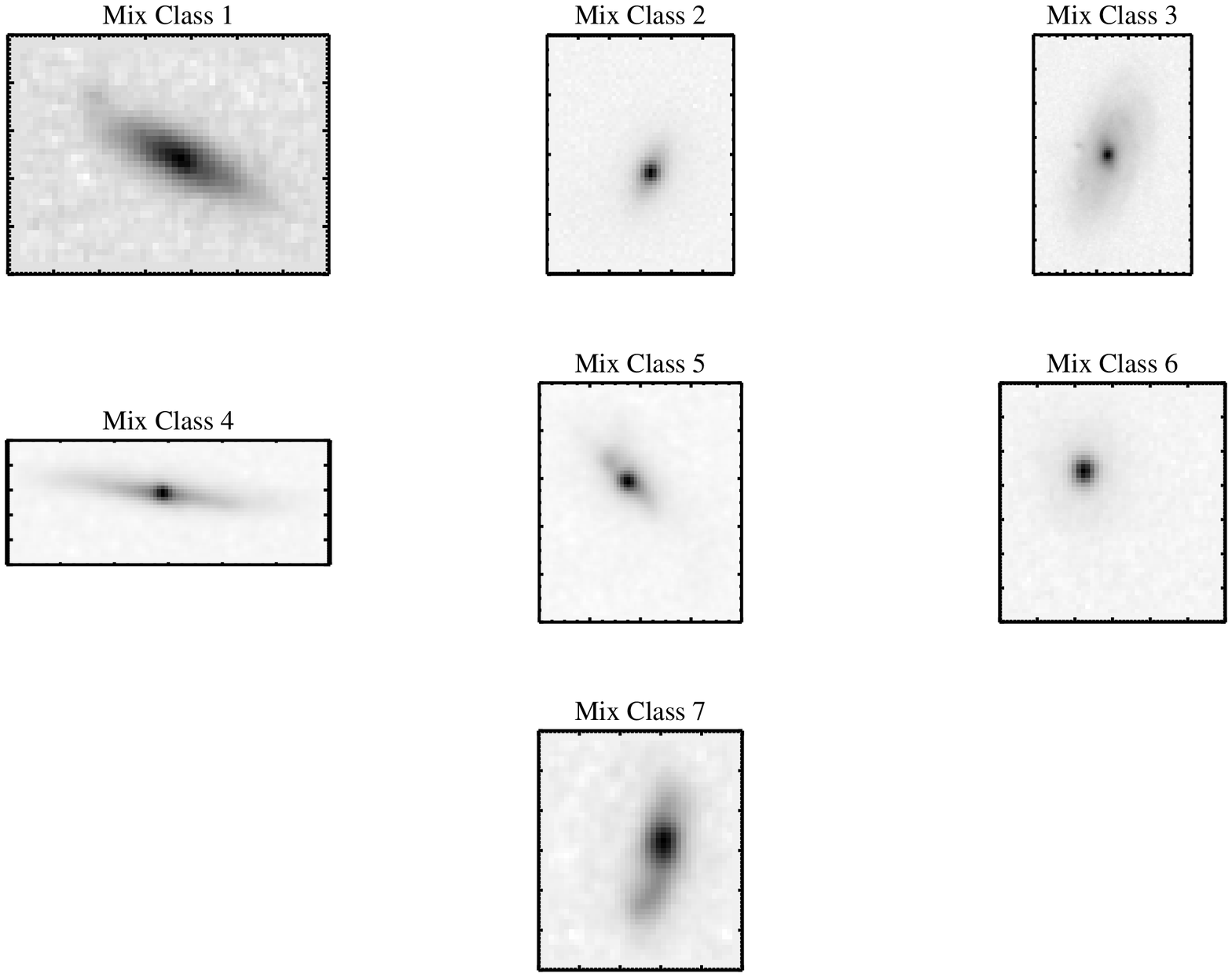}{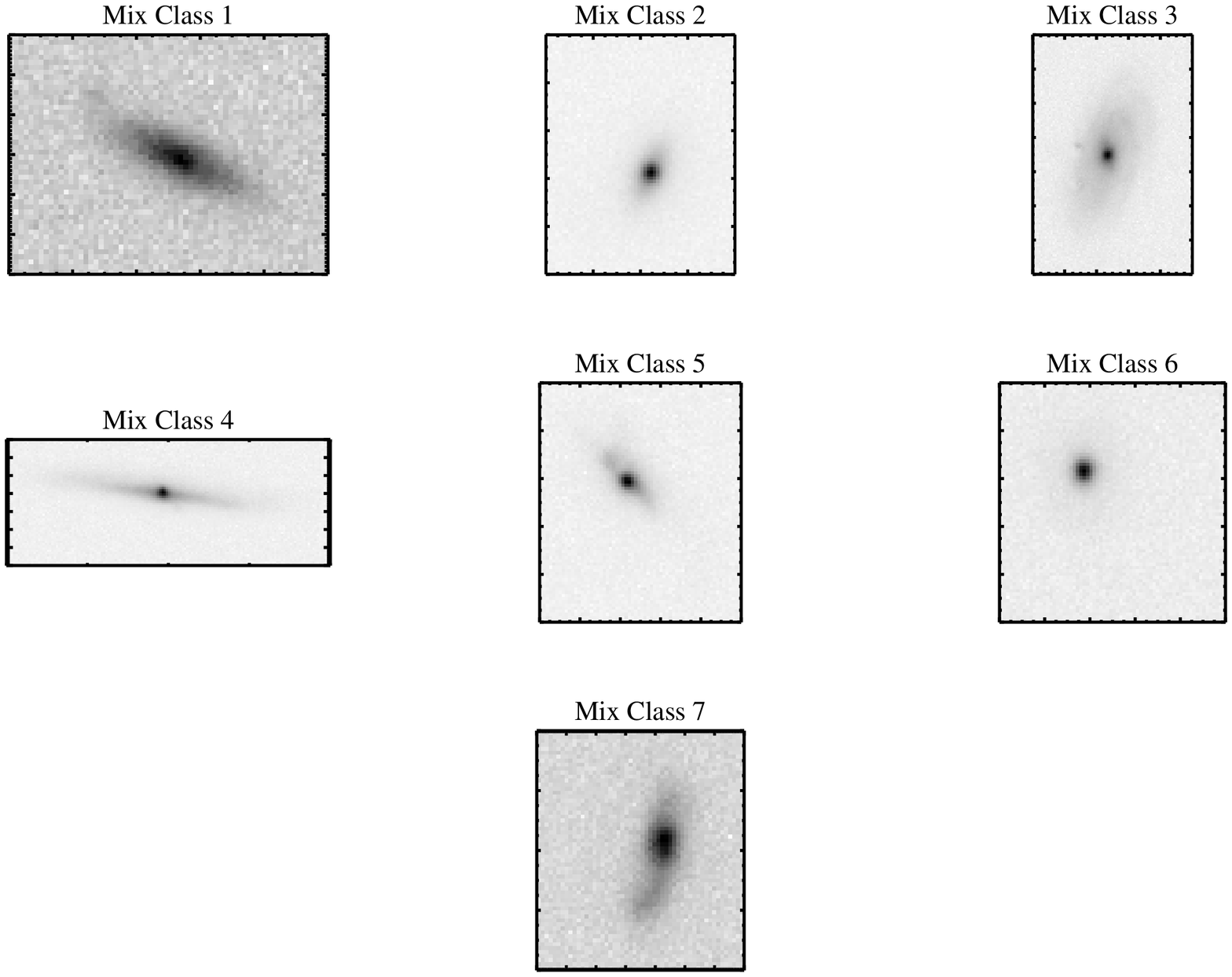}}}
\caption{Galaxies from each $M_k$, shown before artificial redshifting and smoothing to a PSF width of $\beta_0$ (top) and after (bottom). \label{fig16}}
\end{center}
\end{figure}

Because each galaxy has a probability of being in each class, we may calculate mean values of any parameter $z$ (e.g., luminosity) for each class from the equation
\begin{equation}
\langle z_k \rangle = \frac{\sum_{i=1}^{N_{gals}} w_{ik} z_i}{\sum_{i=1}^{N_{gals}} w_{ik}} , 
\label{eq21}
\end{equation}
where $w_{ik}$ is the probability that the $i^{th}$ galaxy is in the $k^{th}$ mixture model class, $z_i$ is the value of the parameter for the $i^{th}$ galaxy, and $N_{gals}$ is the total number of galaxies in the distribution.  We use the normalization $\sum_{k} w_{ik} = 1$, i.e, each galaxy has a probability of one of being in a class.  This assumption may not necessarily be correct, and the addition of a uniform background distribution to the mixture model would likely alleviate this problem; however we did not do this in this analysis and it does not appear that there are any outlier morphologies that would necessitate the addition of a uniform distribution.  The variance of the parameter $z$ may be calculated in a manner similar to the mean.  This is an advantage over classification systems that use discrete classes, such as the Hubble system, as it allows a quantitative comparison between morphological classes and average physical parameters.  We show in Figure \ref{fig14} the images constructed from the mean shapelet coefficients and scales in each class, and in Figure \ref{fig15} we show the mean values of several physical parameters as a function of mixture model class. In addition, we also show a galaxy from each class before and after applying the procedure of \S~\ref{decomp} in Figure \ref{fig16}.  Hereafter, we will denote the $k^{th}$ mixture model class as $M_k$.

\subsection{Description of the Mixture Model Classes}

\label{mix_descript}

Mixture class $M_2$ is the main elliptical class with a probability $\tau_2 \approx 0.266$ (i.e., $M_2$ is expected to contain $\sim 27\%$ of galaxies, see \S~\ref{mixmod}), and $M_6$ a minor elliptical class with $\tau_6 \approx 0.045$.  The association of these classes with early types is evident from the comparison with Hubble types seen in Figure \ref{fig11} and the mean morphologies seen in Figure \ref{fig14}.  The mean Hubble type for $M_2$ and $M_6$ is S0.  The physical properties of $M_2$ and $M_6$ are consistent with early types as well; these galaxies, on average, have higher velocity dispersions, are the reddest, have the smallest scale $\gamma$, are the most concentrated, and have high surface brightness.  Their profiles (i.e., S\'{e}rsic index) are also consistent with early types, as are their spectral eigencoefficients. It should be noted that there is a bias in the computed S\'{e}rsic index towards smaller $n$ (Blanton, private communication).  The effect is small for $n \sim 1$, but becomes larger for $n \sim 3$, such that a value of $n \sim 3$ is more like an actual S\'{e}rsic index of $\sim 4$. The only difference between $M_2$ and $M_6$ is morphological; they have distinctly different coefficients of $p_8$ and $p_9$.  This results in the noticeable twist of the inner isophotes seen in the mean morphology of $M_6$.

Mixture class $M_3$ corresponds to typical late type spirals and has $\tau_3 \approx 0.340$.  The mean morphology of $M_3$ certainly is what we would expect for late spirals, and the mean Hubble type is Sc.  In addition, the physical properties of the galaxies of $M_3$ are consistent with late spirals.  Galaxies of $M_3$ have low velocity dispersions, are blue, large, the least concentrated, have exponential profiles, and low surface brightness.  The spectra of $M_3$ galaxies are also consistent with late types.

The first mixture class ($\tau_1 \approx 0.111$) contains late type spiral galaxies that are viewed edge-on, as is obvious from Figures \ref{fig11} and \ref{fig14}.  These galaxies display many of the same characteristics as those spirals viewed face-on ($M_3$), but are generally dimmer, redder, more concentrated, and differ in their spectra.  These differences are, of course, a result of the viewing angle.  Galaxies viewed edge-on have a higher column density than those seen face-on.  This results in an increase in internal extinction for the galaxies of $M_3$, making them dimmer and redder \citep{elm98}.  Also, this increase in extinction reduces the contribution to the spectrum from the older bulge stars \citep{mis01}.  We have compared the spectral eigencoefficients, $e_j$, for $M_1$ and $M_3$ and find that, on average, the two classes exhibit different $e_j$, with the most noticeable difference in $e_3$.

Mixture class $M_5$ appears to be dominated by early type spirals and ellipticals, with and average Hubble type of Sa and $\tau_5 \approx 0.201$. The projections along the principal components typically show values between the elliptical classes and the spiral classes.  In addition, with the exception of S\'{e}rsic Index, galaxies of $M_5$ typically have physical properties similar to those of $M_2$ and $M_6$.

The fourth mixture class does not appear to be dominated by any particular Hubble type, has $\tau_4 \approx 0.024$, and appears to contain a nearly equal amount of spirals and ellipticals.  Based on the construction of the mean morphology, the average $a_j$, and the value for the S\'{e}rsic index, is seems that $M_4$ has a light profile that falls off steeply near the center and then levels out into an extended halo.  The flux of these galaxies has a tendency to be slightly asymmetric near the center across the minor axis, as seen by comparing $a_4$ and $a_5$.  Galaxies of $M_4$ typically have the largest statistical spread in their physical and morphological properties and show physical properties characteristic of early types.  In addition, $M_4$ is made up of the smallest and closest galaxies, with an average redshift of $z \sim 0.04$, compared to $z \sim 0.055$ for the other mixture classes.  Spirals in $M_4$ are typically small and show a central bulge that is of significantly higher surface brightness and smaller size than the disk.

The last mixture model class, $M_7$, is probably the most interesting.  It is typically made up of late type spirals, both seen edge-on and face-on, with an average Hubble type of Sc and $\tau_7 \approx 0.013$.  The values of $a_3, a_7,$ and $a_8$ are noticeably different from normal late type spirals (i.e., $M_3$), and we interpret their contribution as resulting in a lack of separation between bulge and disk components, or in a bulge with a significantly larger scale length.  In addition, galaxies of $M_7$ appears to typically be slightly asymmetric about the minor axis, and are often asymmetric across the major axis, as evidenced by the large spread in $a_6$.  Mixture class $M_7$ probably contains a significant fraction of morphologically irregular galaxies.  Galaxies of $M_7$ have the highest velocity dispersions and spectral eigenclass, and are the dimmest and bluest.  They are of moderate size, and have central concentrations, S\'{e}rsic indices, and surface brightnesses similar to $M_3$.  In addition, $M_7$ is the only galaxy class with typically negative second and third spectral eigencoefficients, meaning that, on average, they display post-starburst activity in their spectra \citep{yip03}.

\section{CONCLUSIONS AND FUTURE WORK}

\label{conc}

One of the major projects of modern extragalactic astronomy is the pursuit of a quantitative and automatic morphological galaxy classification scheme.  The Hubble sequence is becoming more inadequate, and new systems are needed.  A quantitative description of morphology will allow astronomers to give quantitative relations between a galaxy's morphology and its physical parameters.  In addition, automating the classification scheme will allow the use of large astronomical databases for the analysis of galactic morphology, which will be of great benefit in investigating the physical significance of a galaxy's shape.

In this paper, we have tested the shapelet decomposition method as a quantitative and automatic description of galaxy morphology.  We apply the method to a sample of $\sim 3000$ galaxies from the Sloan Digital Sky Survey and show that galaxies of known Hubble type separate cleanly in shapelet space.  In addition, using the vast amount of SDSS data allows the admission of powerful statistical methods of analysis, such as principal component transforms and the mixture of Gaussians model.  Applying the principal component analysis, we find that the morphological differences of nearby galaxies can be sufficiently described using nine principal components, allowing us to work within a 9-dimensional space spanned by these components.  We show that each principal component contains unique morphological information, and sufficiently separates galaxies that are known to have different morphologies.

Furthermore, we apply a mixture of Gaussians model to describe the density of galaxies in the shapelet space spanned by the principal components, where each Gaussian represents a morphological class.  The mixture model fits the distribution well to seven Gaussians, implying seven classes.  The five `main' classes were shown to correlate well with a broad description of Hubble type, whereas the other two classes were populated by galaxies that were often outliers in shapelet space.  Our method is model-independent, purely objective, and automatic; the fact that the method is able to separate galaxies of different morphologies so well is promising and attests to its efficacy.

We show that galaxies of different morphologies differ, on average, in their physical properties. Much effort has been expended exploring the connection between the morphological appearance of a galaxy and its fundamental physical properties, such as mass, luminosity, and stellar population. There are such connections, and they have been detected through manual classification of galaxies.  Most noteable of these is the correlation between color and Hubble type \citep{rob94}. This paper shows that entirely automated classification schemes can also reveal these trends. It also confirms that connections between morphology and fundamental parameters are loose. Tighter correlations have been observed when comparing manual Hubble classifications to color; however, because of differences in resolution between the sample used in this paper (2.0 kpc) and that used in previous work with Hubble classifications, it is difficult to draw a conclusion as to whether the mixture classes have a closer connection to the spectral properties of galaxies than Hubble types. Indeed, it may well be that the overall morphology of a galaxy is more closely related to accidents of its history than to its fundamental parameters.

The method may be further refined by using data across several wavebands.  We are using SDSS data to decompose the galaxy images in the $u$, $g$, $r$, $i$, and $z$ bands, and applying the principal component analysis to the coefficients in all of these bands (Kelly \& McKay, in preparation).  This will allow us to take into account information of the morphological differences in observing wavelength, and to construct principal components and coefficient arrays that contain spectral information as well.  This will likely serve to make the separation in shapelet space more apparent and may introduce new and interesting morphological classes.  Traditional classification is done on only one observing band, and using information on how a galaxy's shape varies across bands may introduce new information and classifications.  Also, in an attempt to make the coefficient distributions more normally distributed (see \S~\ref{mixclass}), and thus increasing the effectiveness of the mixture of Gaussians model, we are testing the use of shapelets of ellipticity equal to that of the galaxy.  This may take away the axis ratio dependence of the coefficients, which was shown to distort the coefficient distributions such that the coefficients were not normally distributed in the direction of axis ratio information.  It would be interesting to apply the shapelet method to high redshift galaxies and compare results to this local sample.  As we only apply the method to galaxies in the nearby universe ($z < 0.07$), these results are not necessarily applicable for higher redshift galaxies and further analysis of such galaxies is needed.

\section{ACKNOWLEDGMENTS}

B. Kelly acknowledges support from the National Science Foundation REU program, the University of Michigan Center for Theoretical Physics, and the Michigan Space Grant Consortium.  This work was also supported by National Science Foundation grant AST-F007182. We would like to thank Alexandre Refregier for the use of his shapelet code and helpful comments, Michael Blanton for supplying S\'{e}rsic indices and helpful comments, Erin Sheldon and Ben Koester for helping with computer issues, and Sidney van den Bergh, Amy Kimball, and Judith Racusin for helpful comments.  We would also like to thank the referee, Roberto Abraham, for thoughtful comments.

Funding for the creation and distribution of the SDSS Archive has been provided by the Alfred P. Sloan Foundation, the Participating Institutions, the National Aeronautics and Space Administration, the National Science Foundation, the U.S. Department of Energy, the Japanese Monbukagakusho, and the Max Planck Society. The SDSS Web site is {\bf http://www.sdss.org/}.

The SDSS is managed by the Astrophysical Research Consortium (ARC) for the Participating Institutions. The Participating Institutions are The University of Chicago, Fermilab, the Institute for Advanced Study, the Japan Participation Group, The Johns Hopkins University, Los Alamos National Laboratory, the Max-Planck-Institute for Astronomy (MPIA), the Max-Planck-Institute for Astrophysics (MPA), New Mexico State University, University of Pittsburgh, Princeton University, the United States Naval Observatory, and the University of Washington.


\begin{thebibliography}{}

\bibitem[Abazajian et al.(2003)]{aba03} Abazajian, K., et al. 2003, \aj, in press (astro-ph/0305492)
\bibitem[Abraham et al.(1994)]{abr94} Abraham, R.~G., Valdes, F., Yee, H.~K.~C., \& van den Bergh, S.\ 1994, \apj, 432, 75
\bibitem[Abraham et al.(1996a)]{abr96a} Abraham, R.~G., Tanvir, N.~R., Santiago, B.~X., Ellis, R.~S., Glazebrook, K., \& van den Bergh, S.\ 1996, \mnras, 279, L47 
\bibitem[Abraham et al.(1996b)]{abr96b} Abraham, R.~G., van den Bergh, S., Glazebrook, K., Ellis, R.~S., Santiago, B.~X., Surma, P., \& Griffiths, R.~E.\ 1996, \apjs, 107, 1
\bibitem[Abraham \& Merrifield(2000)]{abr00} Abraham, R.~G.~\& Merrifield, M.~R.\ 2000, \aj, 120, 2835
\bibitem[Arp(1966)]{arp66} Arp, H.  1966, \apjs, 14, 1
\bibitem[Ball et al.(2003)]{ball03} Ball, N.M., Loveday, J., Fukugita, M., Nakamura, O., Okamura, S., Brinkmann, J, Brunner, R.J.  2003, submitted to \mnras (astro-ph/0306390)
\bibitem[Blanton et al.(2003a)]{blan03} Blanton, M.R., Hogg, D.W., et al.  2003, \apj, submitted
\bibitem[Blanton et al.(2003b)]{blan03t} Blanton, M.R., Lin, H., Lupton, R.H., Maley, F.M., Young, N., Zehavi, I., and Loveday, J. 2003, AJ, 125, 2276
\bibitem[Chang \& Refregier(2002)]{chang02} Chang, T. \& Refregier, A.  2002, \apj, 570, 447
\bibitem[Connolly et al.(2000)]{con00} Connolly, A.J., Genovese, C., Moore, A.W., Nichol, R.C., Schneider, J., \& Wasserman, L.  2000, \aj, submitted (astro-ph/0008187)
\bibitem[Conselice(2003)]{cons03} Conselice, C.J.  2003, \apjs, submitted (astro-ph/0303065)
\bibitem[de Vaucouleurs(1959)]{devauc59} de Vaucouleurs, G.  1959, Handbuch der Physik, 53, 275 
\bibitem[Doi, Fukugita, \& Okamura(1993)]{doi93} Doi, M., Fukugita, M., \& Okamura, S.\ 1993, \mnras, 264, 832 
\bibitem[Eisenstein(2001)]{eis01} Eisenstein, D.J., et al 2001, AJ, 122, 2267
\bibitem[Elmegreen \& Elmegreen(1982)]{elm82} Elmegreen, D.~M.~\& Elmegreen, B.~G.\ 1982, \mnras, 201, 1021 
\bibitem[Elmegreen(1998)]{elm98} Elmegreen, D.M.  1998, Galaxies \& Galactic Structure (Upper Saddle River: Prentice Hall)
\bibitem[Farge(1992)]{far92} Farge, M.  1992, Ann. Rev. Fluid Mech., 24, 395
\bibitem[Fischer et al.(2000)]{fisc00} Fischer, P., McKay, T.A., Sheldon, E. et al.\ 2000, \aj, 120, 1198 
\bibitem[Frei(1999)]{frei99} Frei, Z.  1999, \apss, 269-270, 577
\bibitem[Fukugita et al.(1996)]{fuk96} Fukugita, M., Ichikawa, T., Gunn, J.E., Doi, M., Shimasaku, K., \& Schneider, D.P.  1996, \aj, 111, 1748
\bibitem[Goderya \& Lolling(2002)]{god02} Goderya, S.N. \&  Lolling, S.M.\ 2002, \apss, 279, 377
\bibitem[Gunn(1998)]{gunn98} Gunn, J.E., Carr, M.A., Rockosi, C.M., Sekiguchi, M., et al. 1998, \aj, 116, 3040
\bibitem[Hastie, Tibshirani, \& Friedman(2001)]{hast01} Hastie, T., Tibshirani, R., \& Friedman, J.  2001, The Elements of Statistical Learning (New York: Springer)
\bibitem[Hogg, Finkbeiner, Schlegel, \& Gunn(2001)]{hogg01} Hogg, D.~W., Finkbeiner, D.~P., Schlegel, D.~J., \& Gunn, J.~E.\ 2001, \aj, 122, 2129 
\bibitem[Hubble(1936)]{hub36} Hubble, E.P.  1936, The Realm of the Nebulae (New Yaven: Yale University Press)
\bibitem[Koopmann \& Kenney(1998)]{koop98} Koopmann, R.~A.~\& Kenney, J.~D.~P.\ 1998, \apjl, 497, L75
\bibitem[Kormendy(1982)]{kor82} Kormendy, J.  1982, in Morphology and Dynamics of Galaxies, Eds. L. Martinet \& M. Mayor (Sauverny: Geneva Observatory)
\bibitem[Massey et al.(2003)]{mas03} Massey, R.J., Refregier, A.R., Conselice, C.J., \& Bacon, D.J.\ 2003, \mnras, submitted (astro-ph/0301449)
\bibitem[Matthews, Morgan, \& Schmidt(1964)]{mat64} Matthews, T.A., Morgan, W.W., \& Schmidt, M.  1964, \apj, 140, 35
\bibitem[Misiriotis, Popescu, Tuffs, \& Kylafis(2001)]{mis01} Misiriotis, A., Popescu, C.~C., Tuffs, R., \& Kylafis, N.~D.\ 2001, \aap, 372, 775 
\bibitem[Moore(1999)]{moo99} Moore, A.W.  1999, in Advances in Neural Information Processing Systems 11, ed. M.~S. Kearns, S.~A. Solla, \& D.~A. Cohn, (Cambridge : MIT Press)
\bibitem[Morgan(1958)]{mor58} Morgan, W.W.  1958, \pasp, 70, 364
\bibitem[Morgan \& Lesh(1965)]{mor65} Morgan, W.W. \& Lesh, J.R.  1965, \apj, 142, 1364
\bibitem[Naim et al.(1995)]{naim95} Naim, A.~et al.\ 1995, \mnras, 274, 1107 
\bibitem[Nakamura et al.(2003)]{nak03} Nakamura, O., Fukugita, M., Yasuda, N., Loveday, J., Brinkmann, J., Schneider, D.~P., Shimasaku, K., \& SubbaRao, M.\ 2003, \aj, 125, 1682 
\bibitem[Odewahn(1995)]{ode95} Odewahn, S.C.  1995, \pasp, 107, 770
\bibitem[Odewahn et al.(2002)]{ode02} Odewahn, S.~C., Cohen, S.~H., Windhorst, R.~A., \& Philip, N.~S.\ 2002, \apj, 568, 539
\bibitem[Pier et al.(2003)]{pier03} Pier, J.~R., Munn, J.~A., Hindsley, R.~B., Hennessy, G.~S., Kent, S.~M., Lupton, R.~H., \& Ivezi{\' c}, {\v Z}.\ 2003, \aj, 125, 1559 
\bibitem[Refregier(2003)]{ref03a} Refregier, A.\ 2003, \mnras, 338, 35 
\bibitem[Refregier \& Bacon(2003)]{ref03b} Refregier, A.~\& Bacon, D.\ 2003, \mnras, 338, 48
\bibitem[Richards(2002)]{ric02} Richards, G. et al.  2002, \aj, 123, 2945
\bibitem[Roberts \& Haynes(1994)]{rob94} Roberts, M.~S.~\& Haynes, M.~P.\ 1994, \araa, 32, 115 
\bibitem[Sandage \& Brucato(1979)]{san79} Sandage, A. \& Brucato, R.  1979, \aj, 84, 472
\bibitem[Schwarz(1979)]{sch79} Schwartz, G.  1979, Annals of Statistics, 6, 461
\bibitem[Smith et al.(2002)]{smith02} Smith, J.~A.~et al.\ 2002, \aj, 123, 2121 
\bibitem[Starck, Donoho, \& Cand{\` e}s(2003)]{star03} Starck, J.~L., Donoho, D.~L., \& Cand{\` e}s, E.~J.\ 2003, \aap, 398, 785 
\bibitem[Stoughton et al.(2002)]{sto02} Stoughton, C.L. et al.  2002, \aj, 123, 485
\bibitem[Strauss et al.(2002)]{str02} Strauss, M.A., et al 2002, AJ, 124, 1810
\bibitem[Thanki, Rhee, \& Lepp(2000)]{tha00} Thanki, S., Rhee, G., \& Lepp, S.\ 2000, American Astronomical Society Meeting, 196
\bibitem[Trinidad(1998)]{tri98} Trinidad, M.~A.\ 1998, Revista Mexicana de Astronomia y Astrofisica Conference Series, 7, 186
\bibitem[van den Bergh, Cohen, \& Crabbe(2001)]{bergh01} van den Bergh, S., Cohen, J.G., \& Crabbe, C.  2001, \aj, 122, 611
\bibitem[van den Bergh(1960)]{bergh60} van den Bergh, S.  1960, \apj, 131, 215
\bibitem[van den Bergh(1976)]{bergh76} van den Bergh, S.  1976, \apj, 206, 883
\bibitem[van den Bergh(1998)]{bergh98} van den Bergh, S.  1998, Galaxy Morphology and Classification (Cambridge, UK: Cambridge University Press)
\bibitem[Yip et al.(2003)]{yip03} Yip, C.W., Connolly, A.J. et. al.  2003, submitted
\bibitem[York et al.(2000)]{york00} York, D. et al.  2000, \aj, 120, 1579

\end{thebibliography}
\end{document}